\documentclass[12pt]{article}
\usepackage{amsfonts}
\usepackage{bm}
\usepackage{amsmath}
\usepackage{epsfig}

\usepackage{multicol}
\usepackage{amssymb,lscape}
\usepackage[matrix,arrow,curve]{xy}

\newcommand{\bmat}{\left(\begin{array}}
\newcommand{\emat}{\end{array}\right)}

\def\gtrsim{\mathrel{\raise.3ex\hbox{$>$\kern-.75em\lower1ex\hbox{$\sim$}}}}

\def\-{\hphantom{-}}

\def\s2{\frac{1}{\sqrt2}}

\def\mg{m_{3/2}}
\def\mg2{m^2_{3/2}}

\def\Dsl{\,\raise.15ex\hbox{/}\mkern-13.5mu D} 

\def\be{\begin{equation}}
\def\ee{\end{equation}}
\def\bea{\begin{eqnarray}}
\def\eea{\end{eqnarray}}

\newcommand{\nn}{\nonumber}

\begin{document}

\pagestyle{plain}

\makeatletter
\@addtoreset{equation}{section}
\makeatother
\renewcommand{\theequation}{\thesection.\arabic{equation}}
\pagestyle{empty}
\begin{center}
\ \

\vskip .5cm

\LARGE{\LARGE\bf Heterotic $\alpha$'-corrections\\ in Double Field Theory \\[10mm]}
\vskip 0.3cm
\large{Oscar A.  Bedoya$^\dag$ ,\  Diego Marqu\'es$^\dag$ and Carmen N\'u\~nez$^{\dag,\S}$
 \\[6mm]}
{\small\it  $^\dag$ Instituto de Astronom\'ia y F\'isica del Espacio (CONICET-UBA)}

 \vskip 0.05cm

 {\small\it  $^\S$ Departamento de F\'isica, FCEyN,
Universidad de Buenos Aires }
  \\[.5 cm]

{\small \verb"{oabedoya , diegomarques , carmen} @iafe.uba.ar"}\\[1cm]

\small{\bf Abstract} \\[0.5cm]\end{center}

{\small We extend the generalized flux formulation
of Double Field Theory to include all  the
 first order
bosonic contributions
to the $\alpha '$ expansion of the
heterotic string low energy effective theory.
The generalized tangent space and duality group are enhanced by $\alpha'$ corrections, and the gauge symmetries are generated by the usual (gauged) generalized Lie derivative in the extended space. The generalized frame receives
derivative corrections  through
the spin connection with torsion, which is incorporated as a new
degree of freedom in the extended bein.
 We compute the generalized fluxes and find
 the Riemann curvature tensor with torsion as one of their components.
All the four-derivative terms of the action, Bianchi identities
and equations of motion are reproduced.
  Using this formalism, we
 obtain the first order $\alpha'$ corrections to the heterotic Buscher rules. The relation of our results to alternative formulations in the literature
is discussed and future research directions are outlined.
}

\newpage
\setcounter{page}{1}
\pagestyle{plain}
\renewcommand{\thefootnote}{\arabic{footnote}}
\setcounter{footnote}{0}

\tableofcontents

\section{Introduction} \label{SECIntro}

The construction of duality invariant formulations of the supergravity limits of string theory has been an active field of research in recent years. A paradigmatic case is Double Field Theory (DFT), where T-duality is incorporated as a manifest symmetry of the universal supergravity sector \cite{Siegel:1993th,Hull:2009mi}. The framework allows to incorporate heterotic vector fields \cite{Siegel:1993th,heteroticHohm}, the Ramond-Ramond fields of type II theories
\cite{TypeII,Dan} and the fermions that complete the supersymmetry multiplets
\cite{Siegel:1993th,Dan,SDFT}. This program led to the full covariantization of supergravities to lowest order in perturbation theory with respect to the T-duality symmetry of string theory. In the process, interesting novel geometric structures emerged, such as the generalized metric \cite{Generalizedmetric} and frame \cite{Siegel:1993th,framelikegeom} including the supergravity fields as components, and a generalized Lie derivative \cite{Siegel:1993th,genLie,Cbracket} that unifies diffeomorphisms and two-form gauge transformations. In this framework, duality invariance is achieved by formally defining the theory on a double space, and the physical space on which supergravity is realized can be recovered upon
enforcing the so-called strong constraint.
 The result is an elegant and powerful reformulation of supergravity in terms of generalized geometric quantities that make T-duality manifest. Interestingly, the duality structure of these theories is manifest even before compactification. For more details and references see \cite{Reviews}.

A natural question is how to incorporate $\alpha'$ corrections in this context. Recently, this question was nicely addressed in \cite{DoubleAlpha}, where a duality invariant CFT
that incorporates $\alpha'$ corrections was presented.
Here we consider the heterotic string, and our goal is to rewrite the massless bosonic sector of the effective low energy theory,
 including all first order
contributions of the $\alpha '$ expansion, in the language of DFT.
This comprises the action, equations of motion, Bianchi identities
and duality transformations.
Although conceptually our approach
looks different from that in \cite{DoubleAlpha}, we
illustrate how both constructions could be connected.

The first  order $\alpha'$ contributions to the
 heterotic string  effective field theory
have an interesting structure. The action
includes  gauge and gravitational Chern-Simons terms in the two-form field strength, in addition to
quadratic terms of the Yang-Mills  field strength  and of the Riemann curvature tensor with torsion.
These contributions were originally obtained
from tree level
 scattering amplitudes of the massless heterotic string states  \cite{cai}.
An alternative method to construct the gravitational part of this action was
developed in \cite{Bergshoeff:1988nn},
 making use of a symmetry that exists between the Yang-Mills and supergravity fields in ten dimensions.
Since this symmetry is an
essential ingredient of our construction, we briefly recall the main idea.

In $d$ dimensional gravity, the spin-connection  plays the role of an $SO(1, d - 1)$
gauge field, that gauges the local Lorentz transformations which are  part of the gauge
symmetries of supergravity. Although this seems to imply that a Riemann
curvature squared action  can be constructed  from the Yang-Mills field strength squared action, simply  replacing
 everywhere  the  gauge connection
by the Lorentz spin connection, these connections
do not have the same behavior under supersymmetry
transformations. However, the replacement of gauge by spin connection
works well
in  the formulation of $d= 10$ supergravity as an
$SO(1, 9)$ Yang-Mills multiplet if
 the  spin connection has  torsion
and the torsion is
proportional to the two-form field strength.
This symmetry between the Yang-Mills
gauge connection and the Lorentz  torsionful
spin connection will be crucial in our formalism, so we will keep it manifest all along the analysis.

Let us start by reviewing the heterotic string low energy effective action
to order $\alpha'$. The massless bosonic degrees of freedom are a $d= 10$ dimensional bein $e_\mu{}^{\bar a}$, a two-form $B_{\mu\nu}$, $n_g = 496$ gauge fields $A_\mu{}^\alpha$ and a dilaton $\phi$, where $\mu,\nu,\dots = 1, \dots, d$ are space-time indices, while $\bar a, \bar b, \dots = 1, \dots, d$ are flat Lorentz indices
and $\alpha,\beta,\dots = 1,\dots,n_g$ are
indices in the adjoint representation of the heterotic gauge group.
The action
can be written as \cite{cai}-\cite{Chemissany:2007he}
\bea
S &=& \int d^{10}x \sqrt{-g}\ e^{-2\phi} \left(R + 4 g^{\mu\nu}\partial_\mu\phi \partial_\nu \phi - \frac 1 {12} g^{\mu\sigma} g^{\nu\tau} g^{\rho\xi}H_{\mu\nu\rho} H_{\sigma\tau\xi} \right.\label{actioneff}\\ && \ \ \ \ \ \ \ \ \ \ \ \ \ \ \ \ \ \ \ \ \ \ \ \ \ \left.- \frac {\alpha'} 4g^{\mu\rho} g^{\nu\sigma} F_{\mu\nu}{}^\alpha F_{\rho \sigma}{}^\beta \kappa_{\alpha\beta} - \frac {\alpha'} 4g^{\mu\rho} g^{\nu\sigma} R^{(-)}_{\mu\nu}{}^\Lambda  R^{(-)}_{\rho \sigma}{}^\Gamma \kappa_{\Lambda\Gamma}\right)\, ,\nn
\eea
where
\bea
H_{\mu\nu\rho} &=& 3\partial_{[\mu} B_{\nu\rho]} - 3 \alpha' \left(\partial_{[\mu} A_\nu{}^\alpha A_{\rho]}{}^\beta \kappa_{\alpha\beta} + \frac 1 3 f_{\alpha\beta\gamma} A_\mu{}^\alpha A_\nu{}^\beta A_\rho{}^\gamma \right)\\
&& \ \ \ \ \ \ \ \ \ \ \ \ - 3\alpha'\left( \partial_{[\mu} \omega^{(-)}_\nu{}^\Lambda  \omega^{(-)}_{\rho]}{}^\Gamma \kappa_{\Lambda\Gamma} + \frac 1 3 f_{\Lambda\Gamma\Sigma}  \omega^{(-)}_\mu{}^\Lambda  \omega^{(-)}_\nu{}^\Gamma \omega^{(-)}_\rho{}^\Sigma\right)
\nn\eea
is the two-form field strength.
As emphasized above, the $\alpha'$ corrections include a Chern-Simons contribution from the gauge fields $A_\mu{}^\alpha$ and a Chern-Simons contribution from the  spin connection with torsion $\omega^{(-)}_\mu{}^\Lambda$. These terms depend on the gauge (Lorentz) Killing metric and structure constants, which are proportional to $\kappa_{\alpha \beta}$ ($\kappa_{\Lambda \Gamma}$) and $f_{\alpha\beta}{}^\gamma$ ($f_{\Lambda\Gamma}{}^\Sigma$) respectively. The indices $\Lambda,\Gamma,\dots = 1,\dots, n_l$ where $n_l = d(d-1)/2$, are adjoint Lorentz indices. We refer to  the Appendix for details on our conventions.  The torsionful spin connection is
\be
\omega^{(-)}_\mu{}^\Lambda (t_{\Lambda})_{\bar a}{}^{\bar b} = \omega_{\mu \bar a}{}^{\bar b}(e) - \frac 1 2 H_{\mu\nu\rho} e_{\bar a}{}^\nu g^{\rho\sigma} e_\sigma{}^{\bar b}\, ,\label{torsionfulspin}
\ee
where $\omega_{\mu \bar a}{}^{\bar b}$ is the usual torsionless spin connection and the two-form field strength plays the role of torsion. Note that since $\omega^{(-)}_\mu{}^\Lambda$ always appears in the action in terms with an $\alpha'$ factor, to $O(\alpha')$ the Chern Simons terms contained in the torsion in (\ref{torsionfulspin}) play no role.
The second line in (\ref{actioneff}) contains the field strengths of the connections
\bea
F_{\mu\nu}{}^\alpha &=& 2 \partial_{[\mu}A_{\nu]}{}^\alpha + f_{\beta\gamma}{}^\alpha A_\mu{}^\beta A_\nu{}^\gamma\, ,\\
 R^{(-)}_{\mu\nu}{}^\Lambda &=&  2 \partial_{[\mu}  \omega^{(-)}_{\nu]}{}^\Lambda + f_{\Gamma\Sigma}{}^\Lambda \omega^{(-)}_\mu{}^\Gamma  \omega^{(-)}_\nu{}^\Sigma\, ,
\eea
the latter being the Riemann tensor defined in terms of the torsionful spin connection.

Written in this form, the  symmetry between the connections is manifest in the action
\bea
A_\mu{}^\alpha  \ \ \ &\leftrightarrow& \ \ \ \omega^{(-)}_\mu{}^\Lambda \ , \ \ \ \ \ \ \ F_{\mu\nu}{}^\alpha \ \ \ \ \leftrightarrow\ \ \ \ R^{(-)}_{\mu\nu}{}^\Lambda \nn\\
\kappa_{\alpha \beta} \ \ \ &\leftrightarrow& \ \ \ \kappa_{\Lambda \Gamma}  \ , \ \ \ \ \ \ \  \ \ \
f_{\alpha \beta}{}^\gamma \ \ \ \ \leftrightarrow\  \ \ \  f_{\Lambda \Gamma}{}^\Sigma\label{symmetry}
\eea
This symmetry extends all along the Bianchi identities (BI). Indeed, the BI for the two-form, gauge and gravitational field strengths read
\bea
\partial_{[\mu}H_{\nu\rho\sigma]} &=& -  \frac 3 4 \alpha' F_{[\mu\nu}{}^\alpha F_{\rho\sigma]}{}^\beta \kappa_{\alpha \beta} - \frac 3 4 \alpha'  R^{(-)}_{[\mu\nu}{}^\Lambda R^{(-)}_{\rho \sigma]}{}^\Gamma \kappa_{\Lambda \Gamma}\, ,\\
 D_{[\mu} F_{\nu\rho]}{}^\alpha &=& \partial_{[\mu} F_{\nu\rho]}{}^\alpha + f_{\beta\gamma}{}^\alpha A_{[\mu}{}^\beta F_{\nu\rho]}{}^\gamma \ =\  0\, ,\\
 D^{(-)}_{[\mu} R^{(-)}_{\nu\rho]}{}^\Lambda &=& \partial_{[\mu} R^{(-)}_{\nu\rho]}{}^\Lambda + f_{\Gamma\Sigma}{}^\Lambda  \omega_{[\mu}^{(-)}{}^\Gamma R^{(-)}_{\nu\rho]}{}^\Sigma \ = \ 0\, .
\eea

At the level of the equations of motion (EOM), the symmetry is more subtle. The reason is that, while the gauge fields $A_\mu{}^\alpha$ are independent degrees of freedom, the torsionful spin connection $\omega^{(-)}_\mu{}^\Lambda$ is not. The
latter depends on the bein, the two-form and the gauge connection, and then a priori there seems to be no reason to consider its EOM.
Let us then begin by writing the well known EOM to $O(\alpha')$ for the dilaton, bein, two-form and gauge fields (see for example \cite{Bergshoeff:1989de}-\cite{Becker:2009zx} and references therein)
\bea
\Delta \phi & =  & R + 4 g^{\mu\nu} (\nabla_\mu  \nabla_\nu \phi  - \partial_\mu \phi \partial_\nu \phi) - \frac 1 {12} g^{\mu\sigma} g^{\nu\tau} g^{\rho\xi}H_{\mu\nu\rho} H_{\sigma\tau\xi} \label{DeltaPhi} \\ && - \frac {\alpha'} 4g^{\mu\rho} g^{\nu\sigma} F_{\mu\nu}{}^\alpha F_{\rho \sigma}{}^\beta \kappa_{\alpha\beta} - \frac {\alpha'} 4g^{\mu\rho} g^{\nu\sigma}  R^{(-)}_{\mu\nu}{}^\Lambda  R^{(-)}_{\rho \sigma}{}^\Gamma \kappa_{\Lambda\Gamma} \ = \ 0\, ,\nn\\
\Delta g_{\mu\nu} &=& R_{\mu\nu} + 2 \nabla_\mu \nabla_\nu \phi - \frac 1 4g^{\sigma \tau} g^{\lambda \xi} H_{\sigma\lambda\mu} H_{\tau\xi\nu}\nn\\
&& - \frac {\alpha'} 2 g^{\sigma \tau } F_{\sigma \mu}{}^\alpha F_{\tau \nu}{}^\beta \kappa_{\alpha \beta}- \frac {\alpha'} 2 g^{\sigma \tau } R^{(-)}_{\sigma \mu}{}^\Lambda  R^{(-)}_{\tau \nu}{}^\Gamma \kappa_{\Lambda \Gamma} \ = \ 0\, , \label{Deltag}\\
\Delta B_{\mu\nu} &=& g^{\rho \sigma} \nabla_\rho \left(e^{-2\phi}H_{\mu\nu\sigma}\right) \ = \ 0\, ,\label{DeltaB}\\
\Delta A_\nu{}^\beta &=& \alpha'g^{\rho \mu } \nabla^{(+,A)}_\rho  \left(e^{-2\phi} F_{\mu\nu}{}^\beta\right) \ = \ 0\, , \label{eqA}
\eea
respectively, where we have defined
\bea
\nabla^{(+,A)}_\rho F_{\mu\nu}{}^\beta &=& \partial_\rho F_{\mu\nu}{}^\beta - \Gamma^{(+)}_{\rho\mu}{}^\sigma F_{\sigma\nu}{}^\beta - \Gamma^{(+)}_{\rho\nu}{}^\sigma F_{\mu\sigma}{}^\beta + f_{\gamma\alpha}{}^\beta A_\rho{}^\gamma F_{\mu\nu}{}^\alpha\, ,\\
\Gamma^{(+)}_{\mu\nu}{}^\rho &=& \Gamma_{\mu\nu}{}^\rho + \frac 1 2 H_{\mu\nu\sigma} g^{\sigma \rho}\, ,\eea
which covariantizes the derivative with respect to ten dimensional diffeomorphisms and gauge transformations. Strictly speaking, the EOM written above are not those that one would get by varying the action with respect to the component fields, but combinations of them.

The heterotic EOM (\ref{DeltaPhi})-(\ref{eqA}) break the symmetry (\ref{symmetry}) because, not being an independent field, there is no EOM
for $\omega_\mu{}^{(-)\Lambda}$.
However,
it is instructive to discuss what happens when varying the action with respect to $\omega_\mu{}^{(-)\Lambda}$,
i.e. treating it as an independent degree of freedom. In this case one obtains the following equation
\be
\Delta \omega^{(-)}_\nu{}^\Gamma\
 = \ \alpha' g^{\rho \mu } \nabla^{(+,-)}_\rho  \left(e^{-2\phi}
R^{(-)}_{\mu\nu}{}^\Gamma\right) \ = \ 0\, , \label{EqOmegaMinus}
\ee
with
\be
 \nabla^{(+,-)}_\rho R^{(-)}_{\mu\nu}{}^\Gamma\ =\ \partial_\rho R^{(-)}_{\mu\nu}{}^\Gamma - \Gamma^{(+)}_{\rho\mu}{}^\sigma R^{(-)}_{\sigma\nu}{}^\Gamma - \Gamma^{(+)}_{\rho\nu}{}^\sigma R^{(-)}_{\mu\sigma}{}^\Gamma + f_{\Lambda\Sigma}{}^\Gamma \omega^{(-)}_\rho{}^\Lambda R^{(-)}_{\mu\nu}{}^\Sigma\, ,\ee
thus restoring the complete symmetry between the gauge and Lorentz sectors, i.e, under (\ref{symmetry}) one has
\be
\Delta A_\nu{}^\beta  \ \ \ \leftrightarrow \ \ \ \Delta \omega^{(-)}_\nu{}^\Gamma\, .
\ee
  Interestingly, it was shown in \cite{Bergshoeff:1989de}
that, to $O(\alpha')$, the equation (\ref{EqOmegaMinus}) is automatically satisfied by the solutions to the other
equations. More precisely, one can show that, to order $\alpha'$, $\Delta \omega^{(-)}_\nu{}^\Gamma $
 can be expressed as a linear combination
of  $\Delta g_{\mu\nu}$ and $\Delta B_{\mu\nu}$, and then it trivially vanishes on-shell.

It is worth mentioning that
 simply replacing $\omega_\mu^{(-)\Lambda}$ by an independent field, say $\tilde \omega_\mu{}^\Lambda$, does not
lead to a first order formulation because, being higher order in derivatives, its equation of motion
(\ref{EqOmegaMinus}) does not provide an algebraic expression relating $\tilde \omega_\mu{}^\Lambda$
with $\omega_\mu^{(-)\Lambda}$. Rather, it is a differential equation that admits
$\omega_\mu^{(-)\Lambda}$ as a solution. However,
due to the lemma proved in \cite{Bergshoeff:1989de}, to $O(\alpha')$
the result of varying the action with respect to the fundamental fields by first
varying the explicit dependence and then adding
the variation through the torsionful spin
connection which implicitly also depends on them,
coincides with the result of
simply considering the explicit variation. We discuss this further in the Appendix.
This suggests that one can still consider a formulation in which $\omega^{(-)}_\mu{}^\Lambda$ is treated
as an independent degree of freedom. This point of view is then useful in order to extend the symmetry between
the gauge and torsionful spin connections to all levels, including the EOM.

In this paper, we encode all these results in the manifestly
T-duality invariant
DFT. Already for the gauge sector this was done in
\cite{Siegel:1993th,heteroticHohm}, where the gauge fields were incorporated in an extended tangent space,
enhancing the $O(d,d)$ duality group
to $O(d, d+n_g)$. Here, we further extend this construction to incorporate the
 gravitational sector to order $\alpha'$, exploiting the above mentioned symmetry between the gauge and
torsionful spin connection. Related constructions can be found in \cite{dualt1}-\cite{CGMW}.

We  work in the generalized flux formulation of DFT \cite{Siegel:1993th,framelikegeom,Exploring}, which is more convenient to display the covariant structures of the effective theory.
In this formulation, the field degrees of freedom appear as components of a generalized frame $E_{\bar A}{}^M$ that parameterizes the quotient $G/H$ (where $G$ is the duality group), and the dilaton is combined with the determinant of the metric in a shifted dilaton. The gauge transformations are encoded in the generalized Lie derivative ${\cal L}$ (to be defined later), which in turn defines  generalized fluxes ${\cal F}_{\bar A \bar B\bar C}$ and ${\cal F}_{\bar A}$. The components of these fluxes contain the covariant quantities of the theory, namely the two-form field strength, the antisymmetrized spin connection, etc. Closure of the gauge algebra imposes constraints which force these fluxes to be covariant under ${\cal L}$, and this leads to a set of closure constraints that take the form of generalized BI.
The fluxes are not covariant under the action of the local subgroup $H$, and then $H$-invariance determines the form of the action up to the closure constraints. The result is an action
quadratic in fluxes, with generalized EOM that can also be written purely in terms of fluxes.

To allow for a description of the
$O(\alpha ')$ corrections to heterotic supergravity
in this formulation, we enlarge
 the  duality group
 to $G = O(d + (d - 1), d + n_g + (d-1)(d-2)/2)$ and  take
 $H = O(1 + (d-1), (d-1) + n_g + (d-1)(d-2)/2) \otimes O(d-1,1) $, so that the dimension of the quotient $G/H$
 allows to accommodate, in addition to the bein and two-form field, $n_g$ gauge  and $n_l=d(d-1)/2$ Lorentz one-form
connections.\footnote{A generalized spin connection was incorporated in an
extended generalized frame
in references \cite{NaturalCurvature}. It would be interesting to explore if this construction is related to the one presented here.} A subgroup of the duality group is gauged and we will argue that a residual $O(d,d)$ global symmetry group is preserved by the gauging.
In this framework, the
generalized Lie derivative reproduces the gauge transformations
of the heterotic fields and the fluxes encode all the covariant building blocks of the theory.
Remarkably, one of the components of the fluxes reduces to the Riemann tensor with torsion upon imposing
the strong constraint, when the  Lorentz connections
included in the generalized bein are identified with the spin connection with torsion.
Moreover, being quadratic in fluxes, the generalized action naturally
 reproduces the Riemann squared term.
In this way, the formalism manages to remove one of the obstructions that impeded
 the inclusion of higher derivative terms in DFT, namely the apparent absence of a
T-duality invariant four-derivative combination built from the
generalized metric
that reduces to the square of the Riemann tensor \cite{Riemann}.

 An interesting application of our formalism is to determine  the
$\alpha '$ corrections to the
Buscher rules of the heterotic massless fields.
These rules play a significant role in the search of
 solutions to the
string equations of motion,  allowing to generate new solutions
from old ones.
Buscher derived the zero slope limit
of the
duality transformations of the fields
from the sigma model worldsheet action  \cite{buscher1}
when there is an isometry (see also \cite{Rocek:1991ps}).
An elegant way to recover these rules is
by performing a canonical transformation
\cite{Alvarez:1994wj}, which shows that the
dual models are
classically equivalent.
The explicit form of the quantum corrections has been pursued
using different methods and some partial results are available
\cite{qbusch,bergortin}.
Here, we obtain
the $O(\alpha')$ corrections to the transformation rules of the massless
heterotic fields in a manifestly duality covariant way. After constructing the generalized metric and transforming it under the factorized T-duality elements of the duality group $G$,
we get the
 explicit results for the $\alpha '$ corrected  duality transformations
of generic background fields. We show how this works for the full $O(d,d,\mathbb{R})$ duality group.

The paper is organized as follows. In Section \ref{SECreview} we briefly review the generalized flux formulation of
DFT and its gauging. We then present the heterotic setup in Section \ref{SEChet}. We
extend the $O(d,d)$ duality group
 to include the extra degrees of freedom that are necessary
to describe the $O(\alpha ')$ corrections to heterotic supergravity, we construct the generalized frame and
study the  gauge transformations of the fields.
The
generalized fluxes are then computed  and  the Bianchi identities they satisfy are found.
The action and  equations of motion are presented in Subsections \ref{SECaction} and \ref{SECeom}, respectively.
In Section \ref{SECgenmet} we construct the generalized metric formulation, and evaluate
the $O(\alpha')$ corrections to the  heterotic Buscher rules.
We also discuss the relation of our formalism with
the double $\alpha'$-geometry introduced in \cite{DoubleAlpha}. In the concluding Section
\ref{SECconclu}, we summarize our results and
outline future directions of research.

\section{Generalized flux formulation of Double Field Theory} \label{SECreview}

Let us begin by briefly reviewing
the generalized flux formulation of  DFT  \cite{Siegel:1993th,framelikegeom,Exploring}. For more details we refer to those references.

The theory is defined on an extended space where derivatives $\partial_M$ span the
fundamental representation of a group $G$. The extended space
indices $M,N,\dots$  take values in the fundamental representation of $G$
and are raised and lowered with the constant and symmetric group metric $\eta_{MN}$. Typically, in order to realize T-duality as a manifest symmetry, the group is taken to be $G = O(d,d)$ and the space is doubled. However, this is not strictly necessary and here instead we will consider a bigger group that contains $O(d,d)$ as a subgroup.

The fields are generalized tensorial densities $T^{M\dots}{}_{N\dots}$ of weight $w(T)$ that transform under {\it generalized diffeomorphisms} as
\bea
({\delta}_\xi T)^{M\dots}{}_{N\dots} = ({\cal L}_\xi T)^{M\dots}{}_{N\dots} &=& \xi^P \partial_P T^{M\dots}{}_{N\dots} + (\partial^M \xi_P - \partial_P \xi^M) \ T^{P\dots}{}_{N\dots} + \dots\nn\\
&&\ \ \ \ \ \ \ \ \ \ \ \ \ \ \ \ \ \, +\ (\partial_N \xi^P - \partial^P \xi_N)\ T^{M\dots}{}_{P\dots} + \dots\nn\\
&&\ \ \ \ \ \ \ \ \ \ \ \ \ \ \ \ \ \, + \ w(T)\ \partial_P \xi^P\  T^{M\dots}{}_{N\dots}\, ,
\eea
where the gauge parameters $\xi^M$ are
generalized vectors themselves  with vanishing weight.

Consider a subgroup $H$ and introduce flat indices $\bar A,\bar B,\dots$ which are acted on by $H$ and are raised and lowered with the constant and symmetric  metric $\eta_{\bar A \bar B}$,
taken to  numerically coincide with $\eta_{MN}$. The elements in $H$ preserve both $\eta_{\bar A \bar B}$, and a symmetric and constant metric $S_{\bar A \bar B}$.

A generalized frame $E_{\bar A}{}^M$ is a basis of generalized vectors of vanishing weight, and can be taken to be parameterized by some of the supergravity field degrees of freedom, namely the metric, two-form, one-form gauge fields, etc. Under generalized diffeomorphisms it transforms as
\be
({\cal L}_\xi E_{\bar A})^M = \xi^P \partial_P E_{\bar A}{}^M + (\partial^M \xi_P - \partial_P \xi^M) E_{\bar A}{}^P\, .
\ee
 The particular parameterization of the generalized frame in terms of the supergravity degrees of freedom depends on the $H$-gauge choice, which we do not need to specify right now. After the action of generalized diffeomorphisms, the gauge choice must be restored. Since it parameterizes the coset $G/H$,  the frame satisfies
\be
E_{\bar A}{}^M \ \eta_{MN} \ E_{\bar B}{}^N = \eta_{\bar A \bar B}\, ,
\ee
and so its inverse is given by $E^{\bar A}{}_M = \eta^{\bar A \bar B} \eta_{MN} E_{\bar B}{}^N$. The dilaton, instead, is contained in a density field $e^{-2d}$, of weight $w(e^{-2d}) = 1$, which transforms as a measure
\be
{\cal L}_\xi e^{-2d} = \partial_P( \xi^P e^{-2d})\, .
\ee

The group of generalized diffeomorphisms closes provided a tower of closure constraints is satisfied. In particular,
the transformation of a tensorial density must be itself a tensorial density
\be
\Delta_{\xi_1} {\cal L}_{\xi_2} T = 0 \ , \ \ \ \ \ \Delta_\xi = \delta_{\xi} - {\cal L}_{\xi}\, ,
\label{closure constraint}
\ee
where ${\cal L}_\xi$ acts on a covariant object, while $\delta_\xi$ faithfully transforms the object. Clearly,  on tensorial densities, one has $\Delta_\xi T = 0 $. Since (\ref{closure constraint}) is not covariant, one should impose the additional constraints that all its gauge transformations vanish as well. The result is a tower of closure constraints that restricts the space of gauge parameters and tensorial densities for which DFT is consistently defined. A stronger constraint,
known as {\it strong constraint} or {\it section condition}, can be imposed
\be
\partial_M  \partial^M \diamond = 0\, ,
\ee
where $\diamond$ represents any combination of fields and gauge parameters. This constraint is sufficient to satisfy the closure constraints (and hence to achieve gauge consistency), but it is not necessary \cite{GDFT,Exploring}. Let us emphasize however that in this paper, for the sake of concreteness and in order to make direct contact with the heterotic supergravity theory in $d=10$-dimensions, we will impose the strong constraint.

The generalized diffeomorphisms  allow to define {\it generalized fluxes}
\bea
{\cal F}_{\bar A \bar B \bar C} &=& ({\cal L}_{E_{\bar A}} E_{\bar B})^M E_{\bar C M}\, ,\\
{\cal F}_{\bar A} &=& e^{-2d} {\cal L}_{E_{\bar A}} e^{2d}\, ,
\eea
which by construction  transform as scalars under generalized diffeomorphisms, up to the closure constraints. When  evaluated on generalized frames, the latter become
\bea
    {\cal Z}_{\bar A\bar B\bar C\bar D} &=& \partial_{[\bar A}{\cal
F}_{\bar
B\bar C\bar D]}-\frac{3}{4}{\cal F}_{[\bar A\bar B}{}^{\bar E} {\cal
F}_{\bar
C\bar D]\bar E}\ =\ 0 \, ,\nn\\
    {\cal Z}_{\bar A \bar B} &=& \partial^{\bar C} {\cal F}_{\bar C\bar
A\bar B}
+2\partial_{[\bar A}{\cal F}_{\bar B]}-{\cal F}^{\bar C} {\cal F}_{\bar
C\bar
A\bar B}\ =\ 0 \, .\label{BIs}
\eea
Moreover, when the strong constraint is enforced, these closure constraints then simply become Bianchi identities.

Since the generalized fluxes are not $H$-covariant, by demanding $H$-invariance the action is fixed to be
\be
	S  = \int dX\ e^{-2  d}\left({\cal F}_{\bar A\bar B\bar C}\ \check{\cal F}^{\bar A \bar B \bar C}  + \ {\cal F}_{\bar A } \check{\cal F}^{\bar A}\right)\, ,\label{action}
\ee
where
\bea
\check{\cal F}^{\bar A \bar B \bar C} &=& {\cal F}_{\bar D\bar E\bar F}\ 	\left[\frac{1}{4} S^{[\bar A|\bar D} \eta^{|\bar B|\bar E} \eta^{|\bar C]\bar F}-\frac{1}{12} S^{\bar A\bar D} S^{\bar B\bar E} S^{\bar C\bar F} - \frac 1 6 \eta^{\bar A\bar D}\eta^{\bar B\bar E}\eta^{\bar C\bar F}\right]\, ,\nn\\
\check{\cal F}^{\bar A} &=& {\cal F}_{\bar B}\left[\vphantom{\frac 1 2}S^{\bar A \bar B} - \eta^{\bar A \bar B}\right]\, .\label{checkedfluxes}
\eea
The action (\ref{action}) is fully invariant under all the global and gauge symmetries, up to the closure constraints (\ref{closure constraint}).

Varying the action with respect to the generalized dilaton and  frame yields the equations of motion
\bea
{\cal G}  &=& 	 (2\partial_{\bar A} - {\cal F}_{\bar A}) \check {\cal F}^{\bar A}	+ {\cal F}_{\bar A\bar B\bar C} \check {\cal F}^{\bar A\bar B\bar C} \ = \ 0 \, ,\label{genEOM}\\ {\cal G}^{\bar A\bar B} 	&=& - 2  \partial^{[\bar A} \check{\cal F}^{\bar B]} + 6({\cal F}_{\bar D} -
\partial_{\bar D}) \check{{\cal F}}^{\bar D[\bar A \bar B]} + 6\check{{\cal F}}^{\bar C \bar D [\bar A} {\cal F}_{\bar C \bar D}{}^{\bar B]} \ = \ 0\, .\nn
\eea

This concludes our brief summary of the gauge symmetries, action, BI and EOM of the generalized flux formulation of DFT.
For more details we refer to the original papers or the reviews \cite{Reviews}.

\subsection{Gauged Double Field Theory}

DFT can be deformed through a gauging procedure \cite{heteroticHohm}, parameterized by an {\it embedding tensor} that satisfies a linear and a quadratic constraint
\be
f_{MNP} = f_{[MNP]} \, , \ \ \ \ \ f_{[MN}{}^P f_{Q]P}{}^R = 0\, ,
\ee
provided (any combination of) the fields and gauge parameters are further restricted to satisfy the constraints
\be
f_{MN}{}^P \partial_P\ \diamond = 0\, .\label{fpartial}
\ee
The embedding tensor dictates how the gauge group is embedded in the global duality group $G$.

Under such a deformation, the generalized diffeomorphisms  become {\it gauged}
\bea
(\widehat {\cal L}_\xi T)^{M\dots}{}_{N\dots} &=& ({\cal L}_\xi
T)^{M\dots}{}_{N\dots} - f_{PQ}{}^M \xi^P T^{Q\dots}{}_{N\dots} +
\dots+ f_{PN}{}^Q \xi^P T^{M\dots}{}_{Q\dots} + \dots \ \ \
\eea
and so do the gauge transformations of the generalized frame and dilaton
\bea
(\widehat {\cal L}_\xi E_{\bar A})^M &=& ({\cal L}_\xi E_{\bar A})^M - f_{PQ}{}^M \xi^P E_{\bar A}{}^Q\, ,\\
\widehat {\cal L}_\xi e^{-2d} &=& {\cal L}_\xi e^{-2d}\, ,
\eea
which in turn induce  gauged contributions to the generalized fluxes
\bea
\widehat {\cal F}_{\bar A \bar B \bar C} &=& {\cal F}_{\bar A \bar B \bar C} - f_{MNP} E_{\bar A}{}^ME_{\bar B}{}^NE_{\bar C}{}^P\, , \\
\widehat{\cal F}_{\bar A} &=& {\cal F}_{\bar A}\, .
\eea

After the gauging procedure, the action, equations of motion, closure constraints, etc. take exactly the same form
as in the previous section, but with hatted fluxes. In this paper we will work with a
gauged DFT (GDFT), but in order to lighten the notation we will  drop the hats. Let us finally comment that this gauging procedure was shown in \cite{GDFT} to be equivalent to a generalized Scherk-Schwarz reduction \cite{Aldazabal:2011nj}.

\section{The Heterotic setup} \label{SEChet}

To accommodate the $O(\alpha ')$ corrections of the heterotic string effective
theory,
we take  the global symmetry group $G = O(d + (d - 1), d + n_g + (d-1)(d-2)/2)$ and consider the subgroup $H = O(1 + (d-1) , (d-1) + n_g + (d-1)(d-2)/2)\otimes O(d-1,1)$. The dimension of the quotient is then
\be {\rm dim}(G/H)  = d^2 + d n_g + d n_l \, ,\ee
which allows to build in a symmetric $d$-dimensional metric $g_{\mu\nu}$, a two-form $B_{\mu\nu}$, $n_g$ one-forms $A_\mu{}^\alpha$ plus other $n_l = d(d-1)/2$ one-forms $\tilde \omega_\mu{}^{\Lambda}$. The indices take values $\mu,\nu\dots = 1,\dots,d$; $\alpha,\beta,\dots = 1,\dots,n_g$ and $ \Lambda, \Sigma, \dots = 1,\dots,n_l$. To make contact with the heterotic string, one has to assume that $d =10$ is the dimension of the physical space-time, $n_g = 496$ is the dimension of the adjoint representation of the $SO(32)$ or $E_8\times E_8$  gauge group, and $n_l=45$ is the dimension of the adjoint representation of the Lorentz group.
In this way, this construction introduces $n_l$ extra connections $\tilde \omega_\mu{}^{ \Lambda}$, which in order to reproduce the heterotic string must be related to the torsionful spin connection $\omega^{(-)}_\mu{}^\Lambda$,
depending on the other fields (see (\ref{torsionfulspin})).

Now, there are two ways to proceed: (a) one treats $\tilde \omega_\mu{}^\Lambda$ as an independent quantity (we have provided evidence in the introduction on why this is possible), or (b) one defines $\tilde \omega_\mu{}^\Lambda$ from the start as a dependent quantity. The second option is subtle in two respects. On the one hand, such dependent quantity must behave properly under duality transformations. More concretely, if one simply replaces $\tilde \omega_\mu{}^\Lambda$ by $\omega_\mu^{(-)\Lambda}$, after a T-duality it will transform to a different quantity that will depend, for instance, on dual derivatives. Then, in order to construct a second order formulation that is well behaved under T-dualities, one must proceed with  caution and consider a quantity that transforms consistently under $O(d,d,\mathbb{R})$ and  reduces to $\omega_\mu^{(-)\Lambda}$ when the strong constraint is solved in the standard space-time coordinates. On the other hand, the DFT formalism enforces an equation of motion for $\tilde\omega_\mu{}^\Lambda$, which must then be trivially satisfied. We will address these issues in due time, and for now just proceed by treating $\tilde \omega_\mu{}^\Lambda$ as an independent quantity.

A generalized vector is of the form $V^M = (V_\mu, V_\alpha, V_\Lambda, V^\mu)$, and the invariant metric in  $G$ is taken to be

\be
\eta_{MN} = \left(\begin{matrix}
0 & 0 & 0 & \delta^\mu{}_\nu \\
0 & \kappa^{\alpha\beta} & 0 & 0\\
0 & 0 & \kappa^{\Lambda\Gamma} & 0\\
\delta_\mu{}^\nu & 0 & 0 &0
\end{matrix}\right)\, ,
\ee
so the co-vector counterpart reads $V_M = (V^\mu, V^\alpha, V^\Lambda, V_\mu)$. Here, $\kappa^{\alpha\beta}$ and $\kappa^{\Lambda\Gamma}$ are proportional to the (inverse) Killing metrics in the adjoint representations of the gauge and Lorentz groups (see the Appendix), with signatures $(0,n_g)$ and $(d-1, (d-1)(d-2)/2)$
respectively.

We now introduce the $H$-invariant metric
\be
S_{\bar A\bar B}
= \left(\begin{matrix}
s^{\bar a \bar b} &  0  & 0 & 0\\
0& \kappa^{\bar \alpha\bar \beta}&0 &0\\
0 & 0& \kappa^{\bar \Lambda \bar \Gamma}& 0\\
0 & 0 & 0&
 s_{\bar a \bar b}
\end{matrix}\right)\, ,\label{S}
\ee
where $s_{\bar a \bar b} = {\rm diag}(-,+,\dots,+)$. Here $\kappa^{\bar \alpha \bar \beta} = e_\alpha{}^{\bar \alpha} \kappa^{\alpha\beta} e_{\beta}{}^{\bar \beta}$ is numerically equivalent to $\kappa^{\alpha\beta}$, which allows to define elements $e_\alpha{}^{\bar \alpha}$ that preserve the
Killing metric of the gauge group, and $\kappa^{\bar \Lambda \bar \Gamma} = e_\Lambda{}^{\bar \Lambda} \kappa^{\Lambda\Gamma} e_{\Gamma}{}^{\bar \Gamma}$ is numerically equivalent to $\kappa^{\Lambda\Gamma}$, which allows to define elements $e_\Lambda{}^{\bar \Lambda}$ that preserve the
Killing metric of the Lorentz group.

\subsection{Generalized frame and gauge transformations}

Consider a generalized $G$-valued frame $E_{\bar A}{}^M $ satisfying
$E_{\bar A}{}^M \eta_{MN} E_{\bar B}{}^N = \eta_{\bar A \bar B}$ with a fixed $H$-gauge choice, and such that it has the following $d$-dimensional   dynamical degrees of freedom:
 a bein $e_\mu{}^{\bar a}$, a two-form $B_{\mu\nu}$, $n_g$ one-forms $A_\mu{}^\alpha$ and $n_l$  one-forms $\tilde \omega_\mu{}^{\Lambda}$. Including also  the
elements $e_\alpha{}^{\bar \alpha}$ and $e_\Lambda{}^{\bar \Lambda}$ introduced above, the frame can be written as
\be
E_{\bar A}{}^M
= \left(\begin{matrix}
e_\mu{}^{\bar a} &  0  & 0 &0\\
\sqrt{\alpha'}A_\mu {}^\beta e_\beta{}^{\bar \alpha}  & e_\alpha{}^{\bar \alpha}
& 0 & 0\\
\sqrt{\alpha '}\tilde \omega_\mu {}^{\Gamma} e_{\Gamma}{}^{\bar \Lambda}  & 0
&e_{\Lambda}{}^{\bar \Lambda}&0\\
-e_{\bar a}{}^\rho c_{\rho\mu}& -\sqrt{\alpha'} e_{\bar a }{}^\rho A_{\rho}{}^\beta \kappa_{\beta\alpha} &
- \sqrt{\alpha'}e_{\bar a}{}^\rho\tilde \omega_{\rho}{}^{\Gamma} \kappa_{\Gamma \Lambda}  &e_{\bar a}{}^{\mu}
\end{matrix}\right)\, ,\label{bein}
\ee
where
\be
c_{\mu\nu} =  B_{\mu\nu} + \frac{\alpha'}{2}
A_\mu{}^\gamma \kappa_{\gamma \beta} A_{\nu}{}^\beta + \frac{\alpha'}{2}\tilde \omega_\mu{}^{\Gamma}
\kappa_{ \Gamma \Sigma}\tilde \omega_{\nu}{}^{\Sigma}\, .
\ee
The fact that such a generalized frame exists globally means that the extended space is generalized paralellizable \cite{Lee:2014mla}.
On the other hand, the dilaton $\phi$ is combined with the determinant of the metric $g$ in the shifted dilaton field
\be
e^{-2d} = \sqrt{-g} e^{-2\phi}\, .
\ee

We now explore the action of generalized diffeomorphisms on the generalized frame and dilaton, and for simplicity
 we impose the section condition and pick the frame in which $\partial_M = (0, 0, 0, \partial_\mu)$. We will assume this for the sake of concreteness in all the rest of the paper, and we will also explicitly
incorporate the $\alpha'$ parameter. The generalized Lie derivative acts as
\bea
({\cal L}_\xi E_{\bar A}){}^M &=& \xi^P \partial_P E_{\bar A}{}^M + (\partial^M \xi_P - \partial_P \xi^M) E_{\bar A}{}^P - \frac{1}{\sqrt{\alpha'}}f_{PQ}{}^M
\xi^P E_{\bar A}{}^Q \, ,\nn\\
{\cal L}_\xi e^{-2d} &=& \partial_P \left(\xi^P e^{-2d}\right) \, ,\label{genLie}
\eea
where the non-vanishing fluxes $f_{PQ}{}^M $ have
only  pure gauge or pure Lorentz indices, thus satisfying the constraint (\ref{fpartial}). Taking  the gauge parameter
 $\xi^M = (\xi_\mu, {\sqrt{\alpha'}}\xi_\alpha, {\sqrt{\alpha'}}\xi_\Lambda, \xi^\mu)$, in components we find
\bea
{\cal L}_\xi \phi &=& \xi^\rho \partial_\rho\phi\, ,\\
{\cal L}_\xi e_\mu{}^{\bar a} &=& \xi^\rho \partial_\rho e_\mu{}^{\bar a} + \partial_\mu \xi^\rho e_\rho{}^{\bar a}\, ,\\
{\cal L}_\xi e_ \alpha{}^{\bar \alpha} &=& \xi^\rho \partial_\rho e_\alpha{}^{\bar \alpha} - f_{\alpha\gamma}{}^{\beta} \xi^\gamma  e_\beta{}^{\bar \alpha} \, ,\\
{\cal L}_\xi A_\mu{}^{\beta} &=& \xi^\rho \partial_\rho A_\mu{}^\beta + \partial_\mu \xi^\rho A_\rho{}^\beta + \partial_\mu \xi^\beta - f_{\alpha \gamma }{}^{\beta} \xi^\alpha A_\mu{}^\gamma\, ,\\
{\cal L}_{\xi} e_\Lambda{}^{\bar \Lambda} &=& \xi^\rho \partial_\rho e_\Lambda{}^{\bar \Lambda} - f_{\Lambda \Sigma}{}^\Gamma \xi^\Sigma e_\Gamma{}^{\bar \Lambda}\, ,\\
{\cal L}_\xi \tilde \omega_\mu{}^{\Gamma} &=& \xi^\rho \partial_\rho \tilde \omega_\mu{}^{\Gamma} + \partial_\mu \xi^\rho \tilde \omega_\rho{}^{\Gamma} + \partial_\mu \xi^{\Gamma}  - f_{\Lambda \Sigma}{}^\Gamma \xi^\Lambda \tilde  \omega_\mu{}^{\Sigma}\, , \label{transftildeomega}\\
{\cal L}_\xi B_{\mu\nu} &=& \xi^\rho \partial_\rho B_{\mu\nu} - 2 \partial_{[\mu}\xi^\rho B_{\nu]\rho} + 2 \partial_{[\mu} \tilde \xi_{\nu]} + \alpha' \partial_{[\mu} A_{\nu]}{}^\alpha \kappa_{\alpha \beta} \xi^\beta + \alpha'\partial_{[\mu} \tilde \omega_{\nu]}{}^\Lambda \kappa_{\Lambda\Gamma} \xi^\Gamma\, ,\ \ \ \ \ \  \ \ \ \ \label{transfB}\eea
where we have defined
\be
\tilde \xi_\mu = \xi_\mu - \frac{\alpha'}{2} \left(A_\mu{}^\alpha \kappa_{\alpha\beta}\xi^\beta + \tilde \omega_\mu{}^\Lambda \kappa_{\Lambda\Gamma}\xi^\Gamma\right)\, .
\ee
The last three terms in (\ref{transfB}) include the gauge and Lorentz transformation of the two-form that implement the Green-Schwarz mechanism \cite{Green:1984sg}. Such a transformation guarantees that the field strength of the two-form, which includes the Chern-Simons terms, is gauge and Lorentz invariant.

The degrees of freedom allowed by the dimension of the quotient $G/H$ suggest that we can take $e_\alpha{}^{\bar \alpha}$ and $e_\Lambda{}^{\bar \Lambda}$ constant. Note however that a generalized diffeomorphism generates a gauge transformation on these quantities, shifting them to non-constant matrices. Then, in order to preserve the gauge choice in which these matrices are constant, a gauge-restoring infinitesimal $H$-transformation  is necessary. Consider $h \in H \in G$, such that
\be
h_{\bar A}{}^{\bar C} \eta_{\bar C \bar D} h_{\bar B}{}^{\bar D} = \eta_{\bar A \bar B} \ , \ \ \ \ \ \ h_{\bar A}{}^{\bar C} S_{\bar C \bar D} h_{\bar B}{}^{\bar D} = S_{\bar A \bar B}\, .
\ee
For $h$ sufficiently close to the identity $h_{\bar A}{}^{\bar B} = \delta_{\bar A}{}^{\bar B} + \Lambda_{\bar A}{}^{\bar B}$, the above conditions impose
\be
\Lambda_{\bar A}{}^{\bar C} \eta_{\bar C\bar B} = - \Lambda_{\bar B}{}^{\bar C} \eta_{\bar C\bar A} \ , \ \ \ \ \ \ \Lambda_{\bar A}{}^{\bar C} S_{\bar C\bar B} = - \Lambda_{\bar B}{}^{\bar C} S_{\bar C\bar A}\, .
\ee
Since $H$ is a symmetry of the theory, one can equivalently define the gauge transformations as
\be
\delta_\xi E_{\bar A}{}^M = {\cal L}_\xi E_{\bar A}{}^M - \Lambda_{\bar A}{}^{\bar B} E_{\bar B}{}^M\, ,
\ee
where the last term is introduced to restore the gauge fixing. It is easy to see that the particular gauge choice $e_{\alpha}{}^{\bar \alpha} = const.$ and $e_{\Lambda}{}^{\bar \Lambda} = const.$ is restored through
\be
\Lambda_{\bar A}{}^{\bar B} = \left(\begin{matrix} 0 & 0 & 0 & 0 \\ 0& \Lambda^{\bar \alpha}{}_{\bar \beta} & 0 & 0 \\ 0 & 0 &  \Lambda^{\bar \Lambda}{}_{\bar \Gamma} & 0 \\ 0 & 0& 0& 0\end{matrix}\right)\, ,
\ee
with
\be
\Lambda^{\bar \alpha}{}_{\bar \beta} = - e_{\bar \beta}{}^\beta f_{\beta\gamma}{}^\alpha
 \xi^\gamma e_\alpha{}^{\bar \alpha}  \ , \ \ \ \ \ \
 \Lambda^{\bar \Lambda}{}_{\bar \Gamma} = - e_{\bar \Gamma}{}^\Gamma f_{\Gamma\Sigma}{}^\Lambda
 \xi^\Sigma e_\Lambda{}^{\bar \Lambda}\, ,
 \ee
which enforces
\be
\delta_\xi e_\alpha{}^{\bar \alpha} = 0 \ , \ \ \ \ \ \ \delta_\xi e_\Lambda{}^{\bar \Lambda} = 0\, ,
\ee
and preserves the form of the other gauge transformations. In particular, $e_\alpha{}^{\bar \alpha}$
 and $e_\Lambda{}^{\bar \Lambda}$ can be taken to coincide with the identity, implying the equivalence between barred and un-barred gauged and Lorentz indices.

 We note that a subgroup of the global symmetry group is gauged by $f_{MN}{}^P$. Since the embedding tensor has only pure gauge and Lorentz components, there is a residual $O(d,d)$ global symmetry that generates the familiar T-duality transformations.

\subsection{Generalized fluxes}\label{SECfluxes}

Given the
generalized frame
and generalized Lie derivative  defined  in (\ref{bein}) and (\ref{genLie}), respectively, we are now ready to compute the generalized fluxes
\bea
{\cal F}_{\bar A \bar B \bar C} &=& ({\cal L}_{E_{\bar A}} E_{\bar B})^M E_{\bar C M}\, ,\\
{\cal F}_{\bar A} &=& e^{-2d} {\cal L}_{E_{\bar A}} e^{2d}\, .
\eea
Using the above parameterization and  imposing the strong constraint, one is left with the following non-vanishing components
\begin{multicols}{3}
\begin{description}
\item ${\cal F}_{\bar a \bar b \bar c}=  - e_{\bar a}{}^\mu
e_{\bar b}{}^\nu e_{\bar c}{}^\rho H_{\mu\nu\rho} \, , $
\item ${\cal F}_{\bar a \bar b}{}^{\bar c} =  2 \omega_{\mu [\bar b}{}^{\bar c} e_{\bar a]}{}^\mu \, , $
\item ${\cal F}_{\bar a \bar b}{}^{\bar \alpha}  =  - e_{\bar a}{}^\mu e_{\bar b}{}^\nu e_\alpha{}^{\bar \alpha} \sqrt{\alpha'} F_{\mu\nu}{}^\alpha  \, ,$
\item $ {\cal F}_{\bar a}{}^{\bar \alpha \bar \beta} =
e_{\bar a}{}^\mu e_\beta{}^{\bar \beta}\kappa^{\alpha\beta} D_\mu e_\alpha{}^{\bar \alpha}\, , $
\item $ {\cal F}_{\bar a} =
 2 e_{\bar a}{}^\mu F_\mu + 2 \omega_{\mu[\bar b}{}^{\bar b}
 e_{\bar a]}{}^\mu \, , $
\item[~~~~~] $ {\cal F}^{\bar \alpha \bar \beta \bar \gamma} = -\frac 1 {\sqrt{\alpha'}} e_\alpha{}^{ \bar \alpha}e_\beta{}^{ \bar \beta}e_\gamma{}^{ \bar \gamma} f^{\alpha \beta\gamma}\, , $
\item[~~~~~]
$ {\cal F}_{\bar a}{}^{\bar \Lambda \bar \Gamma} =
 e_{\bar a}{}^\mu e_\Gamma{}^{\bar \Gamma}\kappa^{\Lambda\Gamma}
\tilde D_\mu e_\Lambda{}^{\bar \Lambda}\, ,$
\item[~~~~~]
${\cal F}_{\bar a \bar b}{}^{\bar \Lambda}  =
 - e_{\bar a}{}^\mu e_{\bar b}{}^\nu e_\Lambda{}^{\bar \Lambda} \sqrt{\alpha'}
 \tilde R_{\mu\nu}{}^\Lambda \, , $
\item[~~~~~]
$ {\cal F}^{\bar \Lambda \bar \Gamma \bar \Sigma}  =
  -\frac 1 {\sqrt{\alpha'}} e_\Lambda{}^{ \bar \Lambda}
e_\Gamma{}^{ \bar \Gamma}e_\Sigma{}^{ \bar \Sigma} f^{\Lambda \Gamma\Sigma}
\, , $
\item[~~~~~~]
\item[~~~~~~]
\item \be\label{fluxcomponents} \ee
\item[~~~~~~]
\item[~~~~~~]
\end{description}
\end{multicols}
\noindent where
\bea
F_\mu &=& \partial_\mu \phi \, ,\label{fsDilaton}\\
H_{\mu\nu\rho} &=& 3\partial_{[\mu} B_{\nu\rho]} - 3 \alpha' \left(\partial_{[\mu} A_\nu{}^\alpha A_{\rho]}{}^\beta \kappa_{\alpha\beta} + \frac 1 3 f_{\alpha\beta\gamma} A_\mu{}^\alpha A_\nu{}^\beta A_\rho{}^\gamma\right)\nn\\
&& \ \ \ \ \ \ \ \ \ \ \ \ - 3 \alpha' \left(\partial_{[\mu}\tilde \omega_\nu{}^\Lambda \tilde \omega_{\rho]}{}^\Gamma \kappa_{\Lambda\Gamma} + \frac 1 3 f_{\Lambda\Gamma\Sigma} \tilde \omega_\mu{}^\Lambda \tilde \omega_\nu{}^\Gamma \tilde\omega_\rho{}^\Sigma\right)\, ,\label{fs2form}\\
e_{\bar a}{}^\mu \omega_{\mu \bar b}{}^{\bar c} &=& \frac 1 2 \left(\tau_{\bar a \bar b}{}^{\bar c} + s_{\bar a \bar d} s^{\bar c \bar e} \tau_{\bar e \bar b}{}^{\bar d} + s_{\bar b \bar d} s^{\bar c\bar e}\tau_{\bar e \bar a}{}^{\bar d}\right)\ , \ \ \ \ \tau_{\bar a \bar b}{}^{\bar c} = 2 e_{[\bar a}{}^\mu \partial_\mu e_{\bar b]}{}^\nu e_\nu{}^{\bar c}\, , \label{antisymSpin} \\
F_{\mu\nu}{}^\alpha &=&  2 \partial_{[\mu}A_{\nu]}{}^\alpha + f_{\beta\gamma}{}^\alpha A_\mu{}^\beta A_\nu{}^\gamma \, ,\label{fsGauge}\\
\tilde R_{\mu\nu}{}^\Lambda &=&  2 \partial_{[\mu}\tilde \omega_{\nu]}{}^\Lambda + f_{\Gamma\Sigma}{}^\Lambda \tilde \omega_\mu{}^\Gamma \tilde \omega_\nu{}^\Sigma \, ,\label{RiemannFlux}\\
D_\mu e_\alpha{}^{\bar \alpha} &=& \partial_\mu e_\alpha{}^{\bar \alpha} + f_{\alpha \beta}{}^\gamma A_\mu{}^\beta e_\gamma{}^{\bar \alpha} \, ,\label{covdergauge}\\
\tilde D_\mu e_\Lambda{}^{\bar \Lambda} &=& \partial_\mu e_\Lambda{}^{\bar \Lambda} + f_{\Lambda \Gamma}{}^\Sigma \tilde \omega_\mu{}^\Gamma e_\Sigma{}^{\bar \Lambda} \, .\label{covderLorentz}
\eea
We then readily identify  all the covariant building blocks of the theory, namely the field-strengths of the dilaton (\ref{fsDilaton}), the two-form (\ref{fs2form}), the bein (\ref{antisymSpin}) (which is the antisymmetrized spin connection), the gauge fields (\ref{fsGauge}), the extra one-forms (\ref{RiemannFlux}) (which is nothing but the Riemann tensor when
 $\tilde \omega_\mu{}^\Gamma$ is identified with the  spin connection) and the covariant derivatives of the gauge and Lorentz beins (\ref{covdergauge}) and (\ref{covderLorentz}). Of course, the last two quantities are just pure gauge as we showed above, so we expect them not to appear in the action. Moreover, since the action is quadratic in fluxes, one can already anticipate the presence of the Riemann squared term induced by $\alpha'$-corrections. Although somehow expected, the fact that the Riemann tensor appears as one of the components of a generalized flux is very interesting. As discussed in \cite{Riemann}, the Riemann tensor is not a component of the generalized Riemann tensor
introduced in \cite{Stringydiffgeom}, nor can it be generated from a combination of derivatives of the generalized metric. Here, the extension of the tangent space permits to accommodate a spin connection, whose field strength is the Riemann tensor, which then appears as a generalized flux component.

For the sake of completion, let us now compute the checked fluxes (\ref{checkedfluxes})
\bea
\check{\cal F}^{\bar A \bar B \bar C} &=& {\cal F}_{\bar D\bar E\bar F}\ 	\left[\frac{1}{4} S^{[\bar A|\bar D} \eta^{|\bar B|\bar E} \eta^{|\bar C]\bar F}-\frac{1}{12} S^{\bar A\bar D} S^{\bar B\bar E} S^{\bar C\bar F} - \frac 1 6 \eta^{\bar A\bar D}\eta^{\bar B\bar E}\eta^{\bar C\bar F}\right]\, ,\\
\check{\cal F}^{\bar A} &=& {\cal F}_{\bar B}\left[\vphantom{\frac 1 2}S^{\bar A \bar B} - \eta^{\bar A \bar B}\right]\, ,
\eea
which are necessary to build the action and derive the equations of motion. Their non-vanishing components read
\bea
\check {\cal F}^{\bar a} \!&=&\!  s^{\bar a \bar b} {\cal F}_{\bar b}\, , \ \ \ \ \ \  \ \ \ \ \ \  \ \ \ \ \ \  \ \ \ \ \ \  \ \ \ \ \ \  \ \ \ \ \ \ \ \ \check {\cal F}^{\bar a \bar b}{}_{\bar \alpha}\ =\ -  \frac 1 {12} s^{\bar a \bar d }s^{\bar b \bar e} \kappa_{\bar \alpha \bar \beta} {\cal F}_{\bar d\bar e}{}^{\bar \beta}\, ,\nn\\
\check {\cal F}_{\bar a} \!&=&\! - {\cal F}_{\bar a} \, , \ \ \ \ \ \  \ \ \ \ \ \ \ \ \  \ \ \ \ \ \  \ \ \ \ \ \  \ \ \ \ \ \  \ \ \ \ \ \  \check{\cal F}^{\bar a}{}_{\bar b \bar \alpha}\ =\ \frac 1 {12} s^{\bar a \bar d}\kappa_{\bar \alpha \bar \beta} {\cal F}_{\bar d \bar b}{}^{\bar \beta}\, ,\nn\\
\check {\cal F}^{\bar a \bar b \bar c} \!&=&\! - \frac 1 {12} s^{\bar a \bar d}s^{\bar b \bar e}s^{\bar c \bar f} {\cal F}_{\bar d \bar e \bar f} \, , \ \ \ \ \ \ \ \  \ \ \ \ \ \  \ \ \ \ \ \ \ \ \ \check {\cal F}_{\bar a\bar b \bar \alpha}\ =\ - \frac 1 {12} \kappa_{\bar \alpha \bar \beta} {\cal F}_{\bar a \bar b}{}^{\bar \beta}\, ,\nn\\
\check {\cal F}^{\bar a \bar b}{}_{\bar c} \!&=& \! \frac 1 6 {\cal F}_{\bar d \bar c}{}^{[\bar a} s^{\bar b]\bar d} - \frac 1 {12} s^{\bar a \bar d} s^{\bar b \bar e} s_{\bar c\bar f} {\cal F}_{\bar d \bar e }{}^{\bar f} \, , \ \ \ \ \ \ \ \   \check {\cal F}^{\bar a \bar b}{}_{\bar \Lambda}\ =\ - \frac 1 {12} s^{\bar a \bar d }s^{\bar b \bar e} \kappa_{\bar \Lambda \bar \Gamma} {\cal F}_{\bar d\bar e}{}^{\bar \Gamma}\, ,\label{checkedcomponents}\\
\check {\cal F}^{\bar a}{}_{\bar b \bar c}\! &=&\! \frac 1 {12} s^{\bar a \bar d}{\cal F}_{\bar d \bar b \bar c} - \frac 1 6 {\cal F}_{\bar b\bar c}{}^{\bar a} \, ,\ \ \ \ \ \  \ \ \ \ \ \  \ \ \ \ \ \ \ \ \  \check{\cal F}^{\bar a}{}_{\bar b \bar \Lambda}\ =\ \frac 1 {12} s^{\bar a \bar d}\kappa_{\bar \Lambda \bar \Gamma} {\cal F}_{\bar d \bar b}{}^{\bar \Gamma}\, ,\nn \\
\check {\cal F}_{\bar a \bar b \bar c}\! &=&\! \frac 1 4 s_{\bar d [\bar a} {\cal F}_{\bar b \bar c]}{}^{\bar d} - \frac 1 6 {\cal F}_{\bar a \bar b \bar c} \, , \ \ \ \ \ \  \ \ \ \ \ \  \ \ \ \ \ \ \ \ \ \ \check {\cal F}_{\bar a\bar b \bar \Lambda}\ =\ - \frac 1 {12} \kappa_{\bar \Lambda \bar \Gamma} {\cal F}_{\bar a \bar b}{}^{\bar \Gamma} \, .\nn
\eea
Note that the fluxes $\check {\cal F}^{\bar a}{}_{\bar \alpha \bar \beta}$ and $\check {\cal F}^{\bar a}{}_{\bar \Lambda \bar \Gamma}$ vanish, signaling the fact that no kinetic term of the gauge and Lorentz beins will appear in the action.
Also, note that the checked fluxes carry the information of the couplings in the action.

\subsection{Generalized Bianchi identities}\label{SECbi}

We have shown that the closure of the algebra of generalized Lie derivatives leads to a set of closure constraints (\ref{BIs}),
that become BI when the strong constraint is enforced. In terms of fluxes, they read
\bea
    {\cal Z}_{\bar A\bar B\bar C\bar D} &=& \partial_{[\bar A}{\cal
F}_{\bar
B\bar C\bar D]}-\frac{3}{4}{\cal F}_{[\bar A\bar B}{}^{\bar E} {\cal
F}_{\bar
C\bar D]\bar E}\ =\ 0\, ,\\
    {\cal Z}_{\bar A \bar B} &=& \partial^{\bar C} {\cal F}_{\bar C\bar
A\bar B}
+2\partial_{[\bar A}{\cal F}_{\bar B]}-{\cal F}^{\bar C} {\cal F}_{\bar
C\bar
A\bar B}\ =\ 0\, . \label{biflux}
\eea

Let us then compute their components to show how they match the BI of the heterotic string. The non-vanishing components  are
\bea
{\cal Z}_{\bar a \bar b} &=& 2 e_{\bar a}{}^\mu e_{\bar b}{}^\nu \left(2\partial_{[\mu} F_{\nu]} + R_{[\mu\nu]}\right)\, , \label{BIdilaton}\\
{\cal Z}_{\bar a \bar b\bar c \bar d} &=& - e_{\bar a}{}^\mu e_{\bar b}{}^\nu e_{\bar c}{}^\rho e_{\bar d}{}^\sigma \left(\partial_{[\mu}H_{\nu\rho\sigma]} + \frac 3 4 \alpha' F_{[\mu\nu}{}^\alpha F_{\rho\sigma]}{}^\beta \kappa_{\alpha \beta} + \frac 3 4 \alpha' \tilde R_{[\mu\nu}{}^\Lambda \tilde R_{\rho \sigma]}{}^\Gamma \kappa_{\Lambda \Gamma} \right)\, ,\label{BIHflux}\\
{\cal Z}_{\bar a\bar b\bar c}{}^{\bar d} &=& \frac 3 4 e_{[\bar a}{}^\mu e_{\bar b|}{}^\nu R_{\mu\nu|\bar c]}{}^{\bar d} \, ,\label{BIRiemann}\\
{\cal Z}_{\bar a \bar b \bar c}{}^{\bar \alpha} &=& - \frac 3 4 e_{\bar a}{}^\mu e_{\bar b}{}^\nu e_{\bar c}{}^\rho e_\alpha{}^{\bar \alpha} {\sqrt{\alpha'}} D_{[\mu} F_{\nu\rho]}{}^\alpha\, , \label{BIgauge}\\
{\cal Z}_{\bar a \bar b \bar c}{}^{\bar \Lambda} &=& - \frac 3 4 e_{\bar a}{}^\mu e_{\bar b}{}^\nu e_{\bar c}{}^\rho e_\Lambda{}^{\bar \Lambda} {\sqrt{\alpha'}} \tilde D_{[\mu} \tilde R_{\nu\rho]}{}^\Lambda \, ,\label{BIRiemann2}\\
{\cal Z}_{\bar a \bar b}{}^{\bar \alpha \bar \beta} &=& e_{\bar a}{}^\mu e_{\bar b}{}^\nu \left(D_{[\mu|} e_\beta{}^{\bar \beta} \kappa^{\alpha\beta} D_{|\nu]}e_\alpha{}^{\bar \alpha} + \frac 1 2 e_\beta{}^{\bar \beta}\kappa^{\alpha\beta} D_{[\mu}D_{\nu]}e_\alpha{}^{\bar \alpha} - \frac 1 4 F_{\mu\nu}{}^\gamma f_\gamma{}^{\alpha\beta} e_\alpha{}^{\bar \alpha} e_\beta{}^{\bar \beta}\right)\\
{\cal Z}_{\bar a \bar b}{}^{\bar \Lambda \bar \Gamma} &=& e_{\bar a}{}^\mu e_{\bar b}{}^\nu \left(\tilde D_{[\mu|} e_\Gamma{}^{\bar \Gamma} \kappa^{\Lambda\Gamma} \tilde D_{|\nu]}e_\Lambda{}^{\bar \Lambda} + \frac 1 2 e_\Gamma{}^{\bar \Gamma}\kappa^{\Lambda\Gamma} \tilde D_{[\mu}\tilde D_{\nu]}e_\Lambda{}^{\bar \Lambda} - \frac 1 4 \tilde R_{\mu\nu}{}^\Sigma f_\Sigma{}^{\Lambda\Gamma} e_\Lambda{}^{\bar \Lambda} e_\Gamma{}^{\bar \Gamma}\right) \ \ \ \ \ \ \ \\
{\cal Z}_{\bar a}{}^{\bar \alpha\bar \beta \bar \gamma} &=& - \frac 3 {4\sqrt{\alpha'}} e_{\bar a}{}^\mu  e_\alpha{}^{\bar \alpha}e_\beta{}^{\bar \beta}e_\gamma{}^{\bar \gamma} A_\mu{}^\epsilon f_{\epsilon\eta}{}^{[\alpha} f^{\beta \gamma]\eta}\, , \\
{\cal Z}_{\bar a}{}^{\bar \Lambda\bar \Gamma \bar \Sigma} &=& - \frac 3 {4\sqrt{\alpha'}} e_{\bar a}{}^\mu  e_\Lambda{}^{\bar \Lambda}e_\Gamma{}^{\bar \Gamma}e_\Sigma{}^{\bar \Sigma} \tilde \omega_\mu{}^\Xi f_{\Xi\Pi}{}^{[\Lambda} f^{\Gamma \Sigma]\Pi} \, ,\\
{\cal Z}^{\bar \alpha\bar \beta \bar \gamma\bar \epsilon} &=& - \frac 3 {4\alpha'} e_\alpha{}^{\bar \alpha}e_\beta{}^{\bar \beta}e_\gamma{}^{\bar \gamma}e_\epsilon{}^{\bar \epsilon} f^{[\alpha\beta}{}_{\delta} f^{\gamma\epsilon]\delta}\, ,\\
{\cal Z}^{\bar \Lambda\bar \Gamma \bar \Sigma\bar \Pi} &=& - \frac 3 {4\alpha'} e_\Lambda{}^{\bar \Lambda}e_\Gamma{}^{\bar \Gamma}e_\Sigma{}^{\bar \Sigma}e_\Pi{}^{\bar \Pi} f^{[\Lambda\Gamma}{}_{\Xi} f^{\Sigma\Pi]\Xi}\, ,
\eea
where
\bea
R_{\mu\nu\bar a}{}^{\bar b} &=& 2\partial_{[\mu} \omega_{\nu]\bar a}{}^{\bar b} - 2\omega_{[\mu|\bar a}{}^{\bar c}\omega_{|\nu]\bar c}{}^{\bar b}\, ,\\
R_{\mu\nu} &=& R_{\rho \mu \bar a}{}^{\bar b} e_{\nu}{}^{\bar a} e_{\bar b}{}^\rho\, ,\\
D_{\mu} F_{\nu\rho}{}^\alpha &=& \partial_\mu F_{\nu\rho}{}^\alpha + f_{\beta\gamma}{}^\alpha A_\mu{}^\beta F_{\nu\rho}{}^\gamma\, ,\\
\tilde D_{\mu} \tilde R_{\nu\rho}{}^\Lambda &=& \partial_\mu \tilde R_{\nu\rho}{}^\Lambda + f_{\Gamma\Sigma}{}^\Lambda \tilde \omega_\mu{}^\Gamma \tilde R_{\nu\rho}{}^\Sigma\, .
\eea

Therefore, we have found the BI for the dilaton field strength and Ricci tensor (\ref{BIdilaton}), the
$\alpha '$ corrected BI for the two-form field strength (\ref{BIHflux}), the first BI for the Riemann tensor (\ref{BIRiemann}), the BI for the gauge field strength (\ref{BIgauge}), the
differential second BI for the torsionful Riemann tensor (\ref{BIRiemann2}), plus other BI  including quadratic constraints that are trivially satisfied by the gauge and Lorentz structure constants.

\subsection{The action}\label{SECaction}

Having computed the
components of the fluxes (\ref{fluxcomponents}) and their checked projections (\ref{checkedcomponents}), it is now
straightforward to compute the action
\be
	S  = \int dX\ e^{-2  d}\left({\cal F}_{\bar A\bar B\bar C}\ \check{\cal F}^{\bar A \bar B \bar C}  + \ {\cal F}_{\bar A } \check{\cal F}^{\bar A}\right)\, .
\ee
In components this reads
\bea
S &=&  \int dX e^{-2d} \left({\cal F}_{\bar a} \check {\cal F}^{\bar a} + {\cal F}_{\bar a \bar b \bar c} \check {\cal F}^{\bar a \bar b \bar c} + 3 {\cal F}_{\bar a \bar b}{}^{\bar c} \check {\cal F}^{\bar a \bar b}{}_{\bar c} + 3 {\cal F}_{\bar a \bar b}{}^{\bar \alpha} \check {\cal F}^{\bar a \bar b}{}_{\bar \alpha} + 3{\cal F}_{\bar a\bar b}{}^{\bar \Lambda} \check {\cal F}^{\bar a \bar b}{}_{\bar \Lambda} \right) \\
&=& \int dX e^{-2d}  \left({\cal F}_{\bar a} s^{\bar a \bar b} {\cal F}_{\bar b} + \frac 1 2 s^{\bar b \bar d } {\cal F}_{\bar a \bar b}{}^{\bar c} {\cal F}_{\bar d \bar c}{}^{\bar a} - \frac 1 4  s^{\bar a \bar d} s^{\bar b \bar e} s_{\bar c \bar f}{\cal F}_{\bar a \bar b}{}^{\bar c} {\cal F}_{\bar d \bar e}{}^{\bar f}- \frac 1 {12} s^{\bar a \bar d} s^{\bar b \bar e} s^{\bar c \bar f}{\cal F}_{\bar a \bar b \bar c} {\cal F}_{\bar d \bar e \bar f}\right.\nn\\
&& \ \ \ \ \ \ \ \ \ \ \ \ \ \ \ \ \ \ \left.-\frac 1 4 s^{\bar a \bar d} s^{\bar b \bar e} \kappa_{\alpha \beta}{\cal F}_{\bar a \bar b}{}^\alpha {\cal F}_{\bar d \bar e}{}^\beta -\frac 1 4 s^{\bar a \bar d} s^{\bar b \bar e} \kappa_{\Lambda \Gamma}{\cal F}_{\bar a \bar b}{}^\Lambda {\cal F}_{\bar d \bar e}{}^\Gamma \right)\, ,\nn
\eea
and after an integration by parts we are left with
\bea
S &=& \int d^{10}x \sqrt{-g}\ e^{-2\phi} \left(R + 4 g^{\mu\nu}\partial_\mu\phi \partial_\nu \phi - \frac 1 {12} g^{\mu\sigma} g^{\nu\tau} g^{\rho\xi}H_{\mu\nu\rho} H_{\sigma\tau\xi} \right.\\ && \ \ \ \ \ \ \ \ \ \ \ \ \ \ \ \ \ \ \ \ \ \ \ \ \ \left.- \frac {\alpha'} 4g^{\mu\rho} g^{\nu\sigma} F_{\mu\nu}{}^\alpha F_{\rho \sigma}{}^\beta \kappa_{\alpha\beta} - \frac {\alpha'} 4g^{\mu\rho} g^{\nu\sigma} \tilde R_{\mu\nu}{}^\Lambda \tilde R_{\rho \sigma}{}^\Gamma \kappa_{\Lambda\Gamma}\right)\, ,\nn
\eea
where $g^{\mu\nu} = e_{\bar a}{}^\mu s^{\bar a \bar b} e_{\bar b}{}^\nu$ and $R = g^{\mu\nu} R_{\mu\nu}$.

This confirms our expectations related to the appearance of the Riemann squared term, and the absence of kinetic terms for the gauge and Lorentz beins. Modulo the identification of $\tilde \omega_\mu{}^\Lambda$ with $\omega^{(-)}_\mu{}^\Lambda$, the action precisely matches (\ref{actioneff}),
the low energy effective action of the heterotic string to order $\alpha'$.

\subsection{Equations of motion}\label{SECeom}
As a final step, we now compute the EOM of the theory. As discussed above, all the EOM are condensed in (\ref{genEOM}), the
generalized EOM
that depend on the generalized fluxes
\bea
{\cal G}  &=& 	 (2\partial_{\bar A} - {\cal F}_{\bar A}) \check {\cal F}^{\bar A}	+ {\cal F}_{\bar A\bar B\bar C} \check {\cal F}^{\bar A\bar B\bar C} \ = \ 0 \, ,\\ {\cal G}^{\bar A\bar B} 	&=& - 2  \partial^{[\bar A} \check{\cal F}^{\bar B]} + 6({\cal F}_{\bar D} -
\partial_{\bar D}) \check{{\cal F}}^{\bar D[\bar A \bar B]} + 6\check{{\cal F}}^{\bar C \bar D [\bar A} {\cal F}_{\bar C \bar D}{}^{\bar B]} \ = \ 0\, .
\eea

The non-vanishing components of these equations are
\bea
{\cal G} &=& R + 4 g^{\mu\nu} (\nabla_\mu  \nabla_\nu \phi  - \partial_\mu \phi \partial_\nu \phi) - \frac 1 {12} g^{\mu\sigma} g^{\nu\tau} g^{\rho\xi}H_{\mu\nu\rho} H_{\sigma\tau\xi} \\ && - \frac {\alpha'} 4g^{\mu\rho} g^{\nu\sigma} F_{\mu\nu}{}^\alpha F_{\rho \sigma}{}^\beta \kappa_{\alpha\beta} - \frac {\alpha'} 4g^{\mu\rho} g^{\nu\sigma} \tilde R_{\mu\nu}{}^\Lambda \tilde R_{\rho \sigma}{}^\Gamma \kappa_{\Lambda\Gamma}\, ,\nn
\eea
where we use the convention
\bea
\nabla_\mu V_\nu &=& \partial_\mu V_\nu - \Gamma_{\mu\nu}{}^\rho V_\rho\, ,\\
\Gamma_{\mu\nu}{}^\rho &=& \omega_{\mu \bar a}{}^{\bar b} e_\nu{}^{\bar a} e_{\bar b}{}^\rho + \partial_\mu e_\nu{}^{\bar a} e_{\bar a}{}^\rho\, ,
\eea
and
\bea
{\cal G}_{\bar a \bar b} &=&  - s_{\bar a \bar c} s_{\bar b \bar d} {\cal G}^{\bar c \bar d} = \frac 1 2 e_{\bar a }{}^\mu e_{\bar b}{}^\nu e^{2\phi}\Delta B_{\mu\nu}\, ,\\
{\cal G}^{\bar a}{}_{\bar b} &=&  e_\rho{}^{\bar a} e_{\bar b}{}^\nu g^{\mu\rho } \Delta g_{\mu\nu}\, ,\\
{\cal G}^{\bar a}{}_{\bar \alpha} &=& - s^{\bar a \bar b} {\cal G}_{\bar b \bar \alpha} = - \frac 1 {2\sqrt{\alpha'}} s^{\bar a \bar b} e_{\bar b}{}^\nu e_{\bar \alpha}{}^\alpha \kappa_{\alpha \beta } e^{2\phi} \Delta A_\nu{}^\beta\, ,\\
{\cal G}^{\bar a}{}_{\bar \Lambda} &=& - s^{\bar a \bar b} {\cal G}_{\bar b \bar \Lambda} = - \frac 1 {2\sqrt{\alpha'}} s^{\bar a \bar b} e_{\bar b}{}^\nu e_{\bar \Lambda}{}^\Lambda \kappa_{\Lambda \Gamma } e^{2\phi} \Delta \tilde \omega_\nu{}^\Gamma\, ,
\eea
where
\bea
\Delta B_{\mu\nu} &=& g^{\rho \sigma} \nabla_\rho \left(e^{-2\phi}H_{\mu\nu\sigma}\right)\, ,\\
\Delta g_{\mu\nu} &=& R_{\mu\nu} + 2 \nabla_\mu \nabla_\nu \phi - \frac 1 4g^{\sigma \tau} g^{\lambda \xi} H_{\sigma\lambda\mu} H_{\tau\xi\nu} \nn\\
&& - \frac {\alpha'} 2 g^{\sigma \tau } F_{\sigma \mu}{}^\alpha F_{\tau \nu}{}^\beta \kappa_{\alpha \beta}- \frac {\alpha'} 2 g^{\sigma \tau } \tilde R_{\sigma \mu}{}^\Lambda \tilde R_{\tau \nu}{}^\Gamma \kappa_{\Lambda \Gamma}\, ,\\
\Delta A_\nu{}^\beta &=& \alpha' g^{\rho \mu } \nabla^{(+,A)}_\rho  \left(e^{-2\phi} F_{\mu\nu}{}^\beta\right)\, ,\\
\Delta \tilde \omega_\nu{}^\Gamma &=& \alpha' g^{\rho \mu } \tilde \nabla^{(+)}_\rho  \left(e^{-2\phi} \tilde R_{\mu\nu}{}^\Gamma\right)\, . \label{ExtraEOM}
\eea
We have defined
\bea
\nabla^{(+,A)}_\rho F_{\mu\nu}{}^\beta &=& \partial_\rho F_{\mu\nu}{}^\beta - \Gamma^{(+)}_{\rho\mu}{}^\sigma F_{\sigma\nu}{}^\beta - \Gamma^{(+)}_{\rho\nu}{}^\sigma F_{\mu\sigma}{}^\beta + f_{\gamma\alpha}{}^\beta A_\rho{}^\gamma F_{\mu\nu}{}^\alpha\, ,\\
\tilde \nabla^{(+)}_\rho \tilde R_{\mu\nu}{}^\Gamma &=& \partial_\rho \tilde R_{\mu\nu}{}^\Gamma - \Gamma^{(+)}_{\rho\mu}{}^\sigma \tilde R_{\sigma\nu}{}^\Gamma - \Gamma^{(+)}_{\rho\nu}{}^\sigma \tilde R_{\mu\sigma}{}^\Gamma + f_{\Sigma\Lambda}{}^\Gamma \tilde \omega_\rho{}^\Sigma \tilde R_{\mu\nu}{}^\Lambda\, ,
\eea
in terms of a torsionful connection
\be
\Gamma^{(+)}_{\mu\nu}{}^\rho = \Gamma_{\mu\nu}{}^\rho + \frac 1 2 H_{\mu\nu\sigma} g^{\sigma \rho}\, .\ee

If the one-form $\tilde \omega_\mu{}^\Lambda$ were identified with the torsionful spin connection $\omega_\mu^{(-)\Lambda}$, one readily  identifies the EOM of the heterotic string as anticipated in Section \ref{SECIntro}.
The last equation (\ref{ExtraEOM}) is the result of varying the action with respect to $\tilde \omega_\nu{}^\Gamma$.
As we discussed before, we expect this equation to admit the torsionful spin connection (\ref{torsionfulspin}) as a solution
\be
\tilde \omega_\mu{}^\Lambda (t_{\Lambda})_{\bar a}{}^{\bar b}= \omega^{(-)}_{\mu\bar a}{}^{\bar b} = \omega_{\mu \bar a}{}^{\bar b}(e) - \frac 1 2 H_{\mu\nu\rho} e_{\bar a}{}^\nu g^{\rho\sigma} e_\sigma{}^{\bar b}\, .\label{torsionfulspin2}
\ee
A well known lemma discussed in \cite{Bergshoeff:1989de}
proves that this is indeed the case. In fact, replacing $\tilde \omega_{\mu}{}^\Gamma = \omega^{(-)}_\mu{}^\Gamma$ in equation (\ref{ExtraEOM}), after some algebra, one can show that\footnote{To derive (\ref{SolExtraEOM}) we have used
\be
R^{(\pm)}_{\bar a \bar b \bar c \bar d} = R^{(\mp)}_{\bar c \bar d \bar a \bar b}  + O(\alpha')\, ,
\ee
and the BI (\ref{BIRiemann2}) which can be rewritten as
\be
\nabla^{(\pm)}_{[\bar c} R^{(\pm)}_{\bar d \bar e]\bar a \bar b} = \pm H_{[\bar c \bar d}{}^{\bar f} R^{(\pm)}_{\bar e]\bar f \bar a \bar b}\, ,
\ee after suitable replacements.}
\be
\alpha' g^{\rho\mu} \nabla^{(+,-)}_\rho\left(e^{-2\phi} R^{(-)}_{\mu\nu\bar a}{}^{\bar b}\right) = \alpha' e^{-2\phi} e_\nu{}^{\bar d} s^{\bar b\bar c} \left(\nabla^{(+)}_{[\bar a} \hat{\cal G}_{\bar c] \bar d} + H_{\bar a \bar c}{}^{\bar e} \hat {\cal G}_{\bar e \bar d} \right)
+ O(\alpha'{}^{2}) \, , \label{SolExtraEOM}
\ee
where
\bea
\hat {\cal G}_{\bar a\bar b} &=&  {\cal G}_{\bar a}{}^{\bar c} s_{\bar c \bar b} - {\cal G}_{\bar a \bar b} = - e_{\bar a }{}^\mu e_{\bar b}{}^\nu\left(\Delta g_{\mu\nu} + \frac 1 2 e^{2\phi} \Delta B_{\mu\nu}\right)\ .
 \eea
 The notation in the covariant derivative in (\ref{SolExtraEOM}) indicates that the curved indices of the torsionful Riemann tensor are covariantized with respect to $\Gamma_{\mu\nu}^{(+)\rho}$ and the flat Lorentz indices are covariantized with respect to $\omega_\mu^{(-)\Lambda}$.

 Equation (\ref{SolExtraEOM}) is expressed in terms of the EOM of the bein and two-form,
so on-shell the extra EOM
(\ref{ExtraEOM}) is satisfied by the solution (\ref{torsionfulspin2}). This means that although we have been treating $\tilde \omega_\mu{}^\Lambda$ as an independent component of the generalized frame from the beginning,  we could have as well considered some dependent quantity (well behaved under T-dualities) that reduces to the torsionful spin connection $\omega^{(-)\Lambda}_\mu$ when the strong constraint is solved in the supergravity frame. In that case, although DFT forces one to consider an extra EOM (\ref{SolExtraEOM}), such equation would be trivially satisfied to $O(\alpha')$. Notice that $\omega_\mu^{(-)\Lambda}$ is a solution of (\ref{SolExtraEOM}) where the strong constraint was already imposed and solved. If one chose a different solution to the strong constraint, then the dependent quantity to be considered would be a T-duality rotation of $\omega^{(-)\Lambda}_\mu$. We will show in the next section that there exists a field-dependent quantity, well behaved under T-dualities, that reduces to $\omega^{(-)\Lambda}_\mu$ when the strong constraint is solved in the standard space-time coordinates.
This will allow us to promote this formulation to a consistent second order formalism.

\section{Generalized metric formulation} \label{SECgenmet}

An alternative formulation of GDFT can be performed in terms of the generalized metric. The
inverse generalized metric is given by
\be
{\cal H}^{MN} = E_{\bar A}{}^M S^{\bar A \bar B} E_{\bar B}{}^N = \left(\begin{matrix} {\cal H}_{\mu\nu} & {\cal H}_{\mu\beta} & {\cal H}_{\mu \Gamma} & {\cal H}_{\mu}{}^{\nu} \\ {\cal H}_{\alpha\nu} & {\cal H}_{\alpha \beta} & {\cal H}_{\alpha \Gamma} & {\cal H}_{\alpha}{}^\nu \\ {\cal H}_{\Lambda\nu} & {\cal H}_{\Lambda\beta} & {\cal H}{}_{\Lambda\Gamma} & {\cal H}_{\Lambda}{}^\nu \\ {\cal H}^\mu{}_\nu & {\cal H}^\mu{}_{\beta} & {\cal H}^\mu{}_{\Gamma} & {\cal H}^{\mu\nu}\end{matrix}\right) \, ,\label{genmet}
\ee
and it is straightforward to compute its components

\bea
{\cal H}^{\mu\nu} &=& g^{\mu\nu} = e_{\bar a}{}^\mu s^{\bar a \bar b} e_{\bar b}{}^\nu \, ,\nn\\
{\cal H}_\mu{}^\nu &=& - c_{\rho \mu} g^{\rho \nu} \, ,\nn\\
{\cal H}_{\mu\nu} &=& g_{\mu\nu} + g^{\rho \sigma } c_{\rho \mu }c_{\sigma \nu} + \alpha' A_\mu{}^\alpha A_\nu{}^\beta \kappa_{\alpha\beta} + \alpha' \tilde \omega_\mu{}^\Lambda \tilde \omega_\nu{}^\Gamma \kappa_{\Lambda \Gamma}\, ,\nn\\
{\cal H}_\alpha{}^\nu &=& - \sqrt{\alpha'} g^{\nu\rho} A_\rho{}^\beta \kappa_{\alpha \beta} \, ,\nn\\
{\cal H}_\Lambda{}^\nu &=& - \sqrt{\alpha'} g^{\nu\rho} \tilde \omega_\rho{}^\Gamma \kappa_{\Lambda \Gamma}\, , \nn\\
{\cal H}_{\mu\beta} &=& \sqrt{\alpha'} \kappa_{\beta \alpha}\left(A_\mu{}^\alpha + g^{\rho \sigma} c_{\rho \mu} A_\sigma{}^\alpha\right)\label{genmetcomponents}\, ,\\
{\cal H}_{\mu\Gamma} &=& \sqrt{\alpha'} \kappa_{\Gamma \Lambda}\left(\tilde \omega_\mu{}^\Lambda + g^{\rho \sigma} c_{\rho \mu} \tilde \omega_\sigma{}^\Lambda\right)\, ,\nn\\
{\cal H}_{\alpha\beta} &=& \kappa_{\alpha \beta} + \alpha' \kappa_{\alpha\eta} \kappa_{\beta \gamma} g^{\rho \sigma} A_\rho{}^\eta A_\sigma{}^\gamma \, ,\nn \\
{\cal H}_{\Lambda\Gamma} &=& \kappa_{\Lambda \Gamma} + \alpha' \kappa_{\Lambda\Sigma} \kappa_{\Gamma \Pi} g^{\rho \sigma} \tilde \omega_\rho{}^\Sigma \tilde \omega_\sigma{}^\Pi \, ,\nn\\
{\cal H}_{\alpha \Gamma} &=& \alpha' \kappa_{\alpha \beta} \kappa_{\Gamma\Lambda } g^{\rho \sigma} A_\rho{}^\beta \tilde \omega_\sigma{}^\Lambda \, .\nn
\eea

The action of GDFT was given
in terms of the generalized metric
in \cite{heteroticHohm} and it has the following form
\bea
S &=& \int dX e^{-2d} \left(\frac 1 8 {\cal H}^{MN} \partial_M {\cal H}^{KL} \partial_N {\cal H}_{KL} - \frac 1 2 {\cal H}^{MN} \partial_N {\cal H}^{KL} \partial_L {\cal H}_{MK}- 2 \partial_M d \partial_N {\cal H}^{MN}\right.\nn\\
&& \ \ \ \ \ \ \ \ \ \ \ \ \ \ \ \ \  \left. + 4 {\cal H}^{MN} \partial_M d \partial_N d - \frac 1 6 f^{MNK}f_{MNK} - \frac 1 4 f^M{}_{NK} f^N{}_{ML} {\cal H}^{KL}\right.\\
&& \ \ \ \ \ \ \ \ \ \ \ \ \ \ \  \ \ \left. - \frac 1 2 f^M{}_{NK}{\cal H}^{NP}{\cal H}^{KQ}\partial_P {\cal H}_{QM} - \frac 1 {12} f^M{}_{KP} f^N{}_{LQ}{\cal H}_{MN}{\cal H}^{KL}{\cal H}^{PQ}\right)\, . \nn
\eea

One can check that this action is equivalent (up to strong constraint violating terms) to (\ref{action}), and one can equally compute the BI and EOM in terms of the generalized metric. Since the  results agree with those obtained in previous sections through the generalized flux formulation, we do not pursue this analysis here.
However, the generalized metric  is more convenient than the
generalized frame formulation to discuss duality symmetries.
This is because the generalized metric is $H$-invariant, and therefore, the action of the duality group $G$ must not be compensated by gauge-fixing $H$-transformations. We make use of this advantage in the following subsection to compute the $\alpha'$-corrections to the heterotic Buscher rules induced by factorized T-dualities.

\subsection{T-duality, $\alpha'$ corrected Buscher rules and $O(d,d,\mathbb{R})$}\label{busch}

We are now in a good position to compute the $\alpha'$ corrections to the Buscher rules, and more generally to discuss the role of the $O(d,d,\mathbb{R})$ symmetry.
In  the absence of $\alpha'$ corrections, the Buscher rules
were derived by Buscher \cite{buscher1} from the sigma model formulation
of  string theory, and they determine how the metric and two-form degrees
of freedom mix $g'(g,B)$ and $B'(g,B)$ under factorized T-dualities. Other
derivations can be found in \cite{Alvarez:1994wj,Rocek:1991ps}
and $\alpha '$ corrections were explored in \cite{qbusch,bergortin},
and references therein.

 Here we apply a different, more direct, strategy. We have seen that the generalized metric takes values in a big duality group $G$, which contains the continuous $O(d,d)$ as a subgroup.
Starting from the generalized metric (\ref{genmet}),
one can then perform an $O(d,d,\mathbb{R})$ rotation that preserves its form. The
dual fields are then extracted from the components of the transformed generalized metric.

Any element of the group $O(d,d,\mathbb{R})$ can be factorized as products of $GL(d)$ transformations, $B$-shifts and factorized T-dualities \cite{Giveon:1994fu}. The first two act trivially on the components of the generalized metric, but the factorized T-dualities require a special treatment.
We have considered so far the space-time indices $\mu,\nu,\dots$, and
we now select a particular direction $z$, such that $\mu = (z,i)$.
A factorized T-duality transformation in the $z$-direction
(not necessarily an isometry in DFT) acts  as
\be
{\cal H}'^{MN} = T_{(z)}^M{}_P {\cal H}^{PQ} T_{(z)}^N{}_Q\, ,
\ee
where
\be
T_{(z)}^M{}_N = \left(\begin{matrix} \delta_\mu^\nu - \delta^z_\mu \delta^\nu_z & 0 & 0 &  \delta_\mu^z \delta_\nu^z \\
0  & \delta_\alpha^\beta & 0 & 0 \\
0& 0 & \delta_\Lambda^\Gamma & 0 \\
\delta_z^\mu \delta_z^\nu & 0 & 0 & \delta^\mu_\nu - \delta^\mu_z \delta_\nu^z \end{matrix}\right) \in G\, .
\ee
leading to the following large system of equations
\bea
{\cal H}'^{ij} &=& {\cal H}^{ij} \ , \ \ \ \ \  {\cal H}'_{ij} \ = \ {\cal H}_{ij}  \ , \ \ \ \ \ {\cal H}'^i{}_\alpha \ = \ {\cal H}^i{}_\alpha  \ , \ \ \ \ \ {\cal H}'^i{}_\Lambda  \ = \  {\cal H}^i{}_\Lambda\, ,\nn\\
{\cal H}'^{iz} &=& {\cal H}^i{}_{z}  \ , \ \ \ \ \ {\cal H}'_{iz}  \ = \ {\cal H}{}_i{}^z  \ , \ \ \ \ \ {\cal H}'^z{}_\alpha \ =  \ {\cal H}_{z\alpha}  \ , \ \ \ \ \ {\cal H}'^z{}_\Lambda  \ = \ {\cal H}_{z\Lambda}\, ,\nn\\
{\cal H}'^{zz} &=& {\cal H}_{zz} \ , \ \ \ \ \  {\cal H}'_{zz}  \ = \  {\cal H}^{zz}  \ , \ \ \ \ \ {\cal H}'_{i\alpha} \ = \ {\cal H}_{i\alpha}  \ , \ \ \ \ \ {\cal H}'_{i\Lambda}  \ = \  {\cal H}_{i\Lambda}\, ,\nn\\
{\cal H}'_{i}{}^j &=& {\cal H}_{i}{}^j \ , \ \ \ \ \  {\cal H}'_{z}{}^z  \ = \  {\cal H}_z{}^z  \ , \ \ \ \ \ {\cal H}'_{z\alpha} \ = \ {\cal H}^{z}{}_\alpha  \ , \ \ \ \ \ {\cal H}'_{z\Lambda}  \ = \  {\cal H}^{z}{}_\Lambda \, ,\nn\\
{\cal H}'_{i}{}^z &=& {\cal H}_{iz}  \ , \ \ \ \ \ {\cal H}'_{z}{} ^j  \ = \  {\cal H}^{zj}  \ , \ \ \ \ \ {\cal H}'_{\alpha\beta} \ = \ {\cal H}_{\alpha\beta} \ , \ \ \ \ \ {\cal H}'_{\Lambda\Gamma}  \ = \  {\cal H}_{\Lambda\Gamma}\, ,\\
{\cal H}'_{\alpha\Lambda} &=& {\cal H}_{\alpha\Lambda}\, ,  \nn
\eea
that can be solved to order $\alpha'$. Once these expressions are evaluated on the particular components of the generalized metric (\ref{genmetcomponents}), to order $\alpha'$ the system admits a unique solution. The computation is long but straightforward, so we  simply state the result. The first order $\alpha'$ corrected Buscher rules are given by
\bea
g'_{zz} &=& \frac 1 {g_{zz}} - \alpha ' \frac{A_z{} \cdot A_z  + \tilde \omega_z \cdot \tilde \omega_z}{g_{zz}^2} \, ,\label{gezz} \\
&& \nn \\
g'_{zi} &=& - \frac{B_{zi}}{g_{zz}} + \frac {\alpha'} 2 \left[\frac {A_z}{g_{zz}} \cdot \left(A_i + \frac {A_z}{g_{zz}}(B_{zi} - g_{zi})\right) + \frac{A_z\cdot A_z}{g_{zz}^2} B_{zi}\right]\\
&& \ \ \ \ \ \ \ \ \   + \frac {\alpha'} 2 \left[\frac {\tilde \omega_z}{g_{zz}} \cdot \left(\tilde \omega_i + \frac {\tilde \omega_z}{g_{zz}}(B_{zi} - g_{zi})\right) + \frac{\tilde\omega_z\cdot \tilde\omega_z}{g_{zz}^2} B_{zi}\right]\, ,\nn\eea
\bea
g'_{ij} &=& g_{ij} - \frac{g_{zi} g_{zj} - B_{zi} B_{zj}}{g_{zz}} - \frac{\alpha'} 2 \left[\frac {A_z}{g_{zz}} \cdot \left(A_i + \frac{A_z}{g_{zz}} (B_{zi} - g_{zi})\right) B_{zj} \right.\\
&& \ \ \ \ \ \ \ \ \ \ \ \ \ \ \ \ \ \ \ \ \ \ \ \ \ \ \ \ \  \ \ \ \ \ \ \ \left.  + \frac {\tilde \omega _z}{g_{zz}} \cdot \left(\tilde\omega_i + \frac{\tilde\omega_z}{g_{zz}} (B_{zi} - g_{zi})\right) B_{zj} + (i \leftrightarrow j) \right]\, ,\nn\eea
\bea
B'_{zi} &=& - \frac{g_{zi}}{g_{zz}} + \frac{\alpha'}{2} \left(\frac {A_z\cdot A_z + \tilde\omega_z \cdot\tilde\omega_z} {g_{zz}^2} g_{zi} - \frac{A_z\cdot A_i + \tilde\omega_z \cdot\tilde\omega_i}{g_{zz}}\right)\, ,\\
&&\nn\\
B'_{ij} &=& B_{ij} - \frac{g_{zi} B_{zj} - B_{zi} g_{zj} }{g_{zz}} + \frac{\alpha'}{2} \left[\frac{A_z\cdot A_z + \tilde\omega_z \cdot\tilde\omega_z}{g_{zz}^2} (B_{iz} g_{jz} - B_{jz} g_{iz}) \right.\label{beij}\\
& &  \ \ \ \ \ \ \ \ \ \ \ \ \ \ \ \ \ \ \ \ \ \ \ \ \ \ \  \ \ \ \ \  \ \ \ \  \ \left. + \frac{A_z}{g_{zz}} \cdot (A_i B_{jz} - A_j B_{iz}) + \frac{\tilde\omega_z}{g_{zz}} \cdot (\tilde\omega_i B_{jz} - \tilde\omega_j B_{iz}) \right] \,
\nn\eea
\bea
A'_z{}^\alpha &=&  - \frac{A_z{}^\alpha}{g_{zz}} + \frac{\alpha' } 2 \frac{A_z \cdot A_z + \tilde\omega_z\cdot \tilde\omega_z}{g_{zz}^2} A_z{}^\alpha\, , \label{Aztransf}\\ &&\nn\\
A'_i{}^\alpha &=&  A_i{}^\alpha + \frac{A_z{}^\alpha}{g_{zz}} (B_{zi} - g_{zi}) - \frac{\alpha' } 2 \left(A_i + \frac{A_z}{g_{zz}} (B_{zi} - g_{zi})\right)\cdot \frac{A_z}{g_{zz}} A_z{}^\alpha \label{Aitransf}\\
&&  \ \ \ \ \ \ \ \ \ \ \ \ \ \ \ \ \ \ \ \ \ \ \ \ \ \ \ \ \ - \frac{\alpha' } 2 \left(\tilde\omega_i + \frac{\tilde\omega_z}{g_{zz}} (B_{zi} - g_{zi})\right)\cdot \frac{\tilde\omega_z}{g_{zz}} A_z{}^\alpha \, ,\nn\eea
where we have used the following notation for the gauge (Lorentz) trace $A_\mu \cdot A_\nu = A_\mu{}^\alpha \kappa_{\alpha\beta} A_\nu{}^\beta$ ($\tilde\omega_\mu \cdot \tilde\omega_\nu = \tilde \omega_\mu{}^\Lambda \kappa_{\Lambda\Gamma} \tilde\omega_\nu{}^\Gamma$). Regarding the dilaton, using that $d' = d$ and the definition of $d$ in terms of the dilaton and the determinant of the metric, one finds
\be
\phi' = \phi - \frac 1 2 \log\left(g_{zz} - \frac{\alpha'} 2 (A_z \cdot A_z + \tilde\omega_z \cdot \tilde\omega_z)\right)\, .
\ee
Finally, due to the symmetry (\ref{symmetry}) between gauge and torsionful gravitational  connections,
one also finds
\bea
\tilde\omega'_z{}^\Lambda &=&  - \frac{\tilde\omega_z{}^\Lambda}{g_{zz}} + \frac{\alpha' } 2 \frac{A_z \cdot A_z + \tilde\omega_z\cdot \tilde\omega_z}{g_{zz}^2} \tilde\omega_z{}^\Lambda\, ,\label{Tduality1form}\\ && \nn\\
\tilde \omega'_i{}^\Lambda &=&  \tilde\omega_i{}^\Lambda + \frac{\tilde\omega_z{}^\Lambda}{g_{zz}} (B_{zi} - g_{zi}) - \frac{\alpha' } 2 \left(A_i + \frac{A_z}{g_{zz}} (B_{zi} - g_{zi})\right)\cdot \frac{A_z}{g_{zz}} \tilde\omega_z{}^\Lambda \label{Tduality1form2}\\
&&  \ \ \ \ \ \ \ \ \ \ \ \ \ \ \ \ \ \ \ \ \ \ \ \ \ \ \ \ \ - \frac{\alpha' } 2 \left(\tilde\omega_i + \frac{\tilde\omega_z}{g_{zz}} (B_{zi} - g_{zi})\right)\cdot \frac{\tilde\omega_z}{g_{zz}} \tilde\omega_z{}^\Lambda \, .\nn
\eea
The one-form gauge and Lorentz fields enter in the action always in $O(\alpha')$ terms. Therefore, the $\alpha'$ corrections to their T-duality transformations (\ref{Aztransf})-(\ref{Aitransf}) and (\ref{Tduality1form})-(\ref{Tduality1form2}) are negligible to $O(\alpha')$. We just included the corrections as they result from the transformation because they might be useful when trying to extend this construction to higher orders.

We have seen that the EOM for $\tilde \omega_\mu{}^\Lambda$ (with the strong constraint solved in the supergravity frame) is solved to $O(\alpha')$ by the torsionful spin connection (\ref{torsionfulspin}), which depends on the bein and
the two-form.  These latter fields have concrete transformation rules under the factorized T-dualities. Then, one must
check explicitly that these transformation rules are consistent with the transformations (\ref{Tduality1form}) and (\ref{Tduality1form2}). This computation was performed in \cite{bergortin} assuming an isometry in the dualized direction, namely the transformation rules of the torsionful spin connection were computed directly from the transformation of its components, and the result is in precise agreement with (\ref{Tduality1form}) and (\ref{Tduality1form2}).

In order to have a genuinely $O(d,d,\mathbb{R})$ invariant formulation, one should not rely on the presence of an isometry. In the general case, since the torsionful spin connection is derivative dependent, after a T-duality it will transform into a dual derivative dependent object. Then, one must find some quantity that, under factorized T-dualities, transforms as in (\ref{Tduality1form})-(\ref{Tduality1form2}) to  $O(\alpha'^0)$, and that reduces to the torsionful spin connection only after implementing and solving the strong constraint in the supergravity frame. We now show that such an object exists, and corresponds to a particular component of the generalized coefficients of anholonomy introduced in \cite{Siegel:1993th},\cite{framelikegeom}.

\subsubsection{T-duality covariant Lorentz connection}

We would like to show that there is a field-dependent object that transforms under factorized T-dualities,
to  $O(\alpha'^0)$, as $\tilde \omega_\mu{}^\Lambda$
\be
\tilde\omega'_z{}^\Lambda =  - \frac{\tilde\omega_z{}^\Lambda}{g_{zz}} \ , \ \ \ \ \
\tilde \omega'_i{}^\Lambda =  \tilde\omega_i{}^\Lambda + \frac{\tilde\omega_z{}^\Lambda}{g_{zz}} (B_{zi} - g_{zi})\ ,  \label{transfotilde}
\ee
and reduces to $\omega_\mu^{(-)\Lambda}$ when the strong constraint is solved in the supergravity frame.

We will now  work in the usual double space, so we consider the invariant $O(d,d,\mathbb{R})$ metric $\eta_{mn}$ with $m,n,\dots = 1,\dots,2d$, and
the generalized double frame ${\cal E}_{\bar m}{}^m$, where the indices $\bar m, \bar n,\dots = 1,\dots, 2d$  are flat $O(1,d-1)^2$ indices. While the $O(d,d,\mathbb{R})$ indices split as $m = (^\mu\ ,\ {}_\mu)$, the flat indices split as $\bar m = (^{\bar a} \ , \ {}_{\bar a})$, and we have
\be
\eta_{mn} = \left(\begin{matrix}0 & \delta^\mu{}_\nu \\ \delta_\mu{}^\nu & 0\end{matrix}\right)\ , \ \ \ \
{\cal E}_{\bar m}{}^m = \left(\begin{matrix}e_\mu{}^{\bar a} & 0 \\ -e_{\bar a}{}^\rho B_{\rho \mu} & e_{\bar a}{}^\mu\end{matrix}\right) \, .
\ee

The generalized frame transforms as follows
\be
{\cal E}'_{\bar m}{}^m = \Lambda_{\bar m}{}^{\bar n} {\cal E}_{\bar n}{}^p T_p{}^m \, ,\label{generaltransf}
\ee
where $T_m{}^n$ is a global element of $O(d,d,\mathbb{R})$, i.e.
\be
T_m{}^p \eta_{pq} T_n{}^q = \eta_{mn}\, ,
\ee
and $\Lambda_{\bar m}{}^{\bar n}$ is a local double Lorentz transformation that satisfies
\be
\Lambda_{\bar m}{}^{\bar p} \eta_{\bar p\bar q} \Lambda_{\bar n}{}^{\bar q} = \eta_{\bar m \bar n} =  \left(\begin{matrix}0 & \delta^{\bar a}{}_{ \bar b} \\ \delta_{\bar a}{}^{ \bar b} & 0 \end{matrix}\right)\ , \ \ \ \ \Lambda_{\bar m}{}^{\bar p} S_{\bar p\bar q} \Lambda_{\bar n}{}^{\bar q} = S_{\bar m \bar n} = \left(\begin{matrix}s^{\bar a \bar b} & 0 \\ 0 & s_{\bar a \bar b} \end{matrix}\right)
\ee
with $s_{\bar a \bar b}$ the Minkowski metric.

We  now explore how the components of the generalized frame transform under $O(d,d,\mathbb{R})$. The elements of this group factorize in $GL(d)$ transformations, $B$-shifts, and factorized T-dualities. The first two preserve the triangular form of the generalized frame, but the latter do not. Then, in order to restore the gauge one has to compensate with a local double Lorentz transformation. A factorized T-duality in the $z$  direction
has the form
\be
T^{(z)}_m{}^n = \left(\begin{matrix}\delta_\mu^\nu - \delta_\mu^z \delta^\nu_z & \delta^\mu_z \delta^\nu_z \\ \delta_\mu^z \delta_\nu^z & \delta^\mu_\nu - \delta^\mu_z\delta_\nu^z\end{matrix}\right)\, .
\ee
Acting with this element on the space-time index of the generalized frame takes it away from the triangular parameterization. However, compensating with the following double Lorentz transformation
\be
\Lambda_{\bar m}{}^{\bar n} = \left(\begin{matrix}\delta_{\bar b}^{\bar a} - s_{\bar b\bar c} \frac{e_z{}^{\bar c} e_z{}^{\bar a}}{g_{zz}} & \frac{e_z{}^{\bar a} e_z{}^{\bar b}}{g_{zz}} \\  s_{\bar a\bar c} s_{\bar b\bar d} \frac{e_z{}^{\bar c} e_z{}^{\bar d}}{g_{zz}} & \delta_{\bar a}^{\bar b} - s_{\bar a\bar c} \frac{e_z{}^{\bar c} e_z{}^{\bar b}}{g_{zz}}\end{matrix}\right)\, ,
\ee
one finds the following transformation rules for the components of the generalized frame
\bea
e'_z{}^{\bar a}&=& \frac {e_z{}^{\bar a}}{g_{zz}} \ , \ \ \ \ e'_i{}^{\bar a} \ =\  e_i{}^{\bar a} - \frac{e_z{}^{\bar a}}{g_{zz}} (g_{iz} - B_{iz}) \, ,\\
e'_{\bar a}{}^z &=& e_{\bar a}{}^z g_{zz} + e_{\bar a}{}^i (g_{iz} - B_{iz}) \ , \ \ \ \ e'_{\bar a}{}^i \ = \  e_{\bar a}{}^i \, ,\\
B'_{zi} &=& - \frac{g_{zi}}{g_{zz}} \ , \ \ \ \ B'_{ij} \ = \  B_{ij} - \frac{g_{zi} B_{zj} - B_{zi} g_{zj} }{g_{zz}}\, .
\eea
The transformations for the bein are only defined up to the diagonal part of the double Lorentz group. One can check that they reproduce the transformations
(\ref{gezz})-(\ref{beij}) for the metric and two form to lowest order in $\alpha'$.

On the other hand, one can define the $O(d,d)$ generalized fluxes (also called generalized coefficients of anholonomy)  in terms of the generalized frame
\be
{\cal F}_{\bar m \bar n \bar p} = {\cal E}_{[\bar m|}{}^m \partial_m {\cal E}_{|\bar n}{}^n {\cal E}_{\bar p]n}\, .
\ee
The  components of these fluxes are detailed in \cite{Exploring}. Under a global T-duality these objects are manifestly invariant, but after the compensating double Lorentz transformation they transform in a non-trivial way \cite{Exploring}
\be
{\cal F}'_{\bar m \bar n \bar p} = 3 \Lambda_{[\bar m|}{}^{\bar q} \partial_{\bar q} \Lambda_{|\bar n}{}^{\bar r} \Lambda_{\bar p]\bar r} + \Lambda_{\bar m}{}^{\bar q}\Lambda_{\bar n}{}^{\bar r}\Lambda_{\bar p}{}^{\bar s} {\cal F}_{\bar q \bar r \bar s}\, .
\ee

To understand the impact of these transformations, it is convenient to perform an $SO(2, 2d-2)$ rotation on flat indices through the element
\be
{\cal O}_{\bar m}{}^{\bar n} = \frac{1}{\sqrt 2} \left(\begin{matrix}\delta^{\bar a}_{\bar b} & - s^{\bar a \bar b} \\ s_{\bar a \bar b} & \delta^{\bar b}_{\bar a}\end{matrix}\right) \ .
\ee
This rotation connects the frame formalism in \cite{Siegel:1993th} and \cite{framelikegeom} with that in \cite{Exploring}. Under this rotation, the
quantities defined above
become
\bea
\hat \eta_{\bar m \bar n} &=& {\cal O}_{\bar m}{}^{\bar p} {\cal O}_{\bar n}{}^{\bar q} \eta_{\bar p \bar q} = \left(\begin{matrix} - s^{\bar a \bar b} & 0 \\ 0 & s_{\bar a \bar b}\end{matrix}\right) \ , \\
\hat S_{\bar m \bar n} &=& {\cal O}_{\bar m}{}^{\bar p} {\cal O}_{\bar n}{}^{\bar q} S_{\bar p \bar q} = \left(\begin{matrix}  s^{\bar a \bar b} & 0 \\ 0 & s_{\bar a \bar b}\end{matrix}\right)\ , \\
\hat \Lambda_{\bar m}{}^{\bar n} &=& {\cal O}_{\bar m}{}^{\bar p} \Lambda_{\bar p}{}^{\bar q} ({\cal O}^{-1})_{\bar q}{}^{\bar n} = \left(\begin{matrix} \delta^{\bar a}_{\bar b} - 2 s_{\bar b \bar c} \frac{e_z{}^{\bar c} e_z{}^{\bar a}}{g_{zz}} & 0 \\ 0 & \delta^{\bar b}_{\bar a} \end{matrix}\right)\, ,\label{DoubleLorentz}
\eea
and the rotated generalized coefficients of anholonomy are given by
\be
\hat {\cal F}_{\bar m \bar n \bar p} = {\cal O}_{\bar m}{}^{\bar q}{\cal O}_{\bar n}{}^{\bar r}{\cal O}_{\bar p}{}^{\bar s} {\cal F}_{\bar q \bar r \bar s}\, .
\ee
Now, they transform under hatted compensating double Lorentz transformations after a factorized T-duality as
\be
\hat {\cal F}'_{\bar m \bar n \bar p} = 3 \hat \Lambda_{[\bar m|}{}^{\bar q} \hat \partial_{\bar q} \hat \Lambda_{|\bar n}{}^{\bar r} \hat \Lambda_{\bar p]\bar r} + \hat \Lambda_{\bar m}{}^{\bar q}\hat \Lambda_{\bar n}{}^{\bar r}\hat \Lambda_{\bar p}{}^{\bar s} \hat {\cal F}_{\bar q \bar r \bar s}\, .
\ee

Let us  pay particular attention to the component $\hat {\cal F}^{\bar a}{}_{\bar b \bar c}$. Under a compensating double Lorentz transformation, it varies as
\be
\hat {\cal F}'^{\bar a}{}_{ \bar b \bar c} = \hat \Lambda^{\bar a \bar q} \hat \partial_{\bar q} \hat \Lambda_{ \bar b}{}^{ \bar r} \hat \Lambda_{\bar c \bar r} + \hat \Lambda_{\bar  b}{}^{\bar q} \hat \partial_{\bar q} \hat \Lambda_ { \bar c}{}^{ \bar r} \hat \Lambda^{\bar a}{}_{\bar r} + \hat \Lambda_{\bar  c}{}^{ \bar q} \hat \partial_{\bar q} \hat \Lambda^{ \bar a \bar r} \hat \Lambda_{\bar b \bar r} + \hat \Lambda^{\bar a \bar q}\hat \Lambda_ {\bar b}{}^{ \bar r}\hat \Lambda_{\bar c}{}^{ \bar s} \hat {\cal F}_{\bar q \bar r \bar s}\, .
\ee
Replacing (\ref{DoubleLorentz}), we see that the first three terms cancel, and one is left with
\be
\hat {\cal F}'^{\bar a}{}_{ \bar b \bar c} = \left(\delta^{\bar a}_{\bar d} - 2 s_{\bar d \bar e} \frac {e_z{}^{\bar e} e_z{}^{\bar a}}{g_{zz}}\right) \hat {\cal F}^{\bar d}{}_ {\bar b \bar c}\, .
\ee
Then, if we define
\be
\tilde \omega_{\mu \bar b}{}^{\bar c} = - \frac 1{\sqrt 2} e_\mu{}^{\bar a} s_{\bar a \bar d} s^{\bar c \bar e} \hat {\cal F}^{\bar d}{}_{\bar b \bar e} \, ,\label{defomegatilde}
\ee
we find that under a factorized T-duality it transforms as
\be
\tilde\omega'_{z \bar b}{}^{\bar c} =  - \frac{\tilde\omega_{z\bar b}{}^{\bar c}}{g_{zz}} \ , \ \ \ \ \
\tilde \omega'_{i\bar b}{}^{\bar c} =  \tilde\omega_{i\bar b}{}^{\bar c} + \frac{\tilde\omega_{z\bar b}{}^{\bar c}}{g_{zz}} (B_{zi} - g_{zi})\ .  \label{transfotilde}
\ee
Moreover, one can check \cite{framelikegeom} that under the definition (\ref{defomegatilde}), $\tilde \omega_{\mu \bar b}{}^{\bar c}$ exactly reduces to $\omega^{(-)}_{\mu \bar b}{}^{\bar c}$ defined in (\ref{torsionfulspin}), when the strong constraint is solved in the supergravity frame.

\subsection{Comparison with Double $\alpha'$-Geometry} \label{SECdoublealpha}

Having computed the generalized metric, it is instructive to compare our approach with that of the double $\alpha'$-geometry presented in \cite{DoubleAlpha}. There, it was realized
that $\alpha'$ corrections  can be obtained from a duality covariant CFT construction. In that approach, the generalized Lie derivative receives an $\alpha'$ correction, T-dualities are not corrected, and the tangent space is the usual double tangent space. In contrast, here we preserve the form of the generalized Lie derivative and extend the duality group by enhancing the generalized tangent space. It is then natural to ask if these two seemingly different approaches can be reconciled.

In \cite{DoubleAlpha}, both the inner product and the generalized Lie derivative receive higher derivative corrections (we  introduce $\alpha'$ explicitly to make the comparison with our results clearer)
\bea
\langle\xi, V\rangle &=& \xi^m \eta_{mn} V^n - \alpha' \underline{\partial_m \xi^n \partial_n V^m}\, ,
\label{z1}\\
 \left({\cal L}_\xi V\right)^m &=& \xi^p \partial_p V^m + (\eta^{mn} \eta_{pq}\partial_n \xi^q - \partial_p  \xi^m) V^p - \alpha' \underline{\eta^{mn}\partial_p V^q \partial_n \partial_q\xi^p} \, .\label{z2}\eea
Here, we use the same convention of the previous subsection, namely  $m,n,\dots = 1,\dots,2d$ are $O(d,d)$ indices, which are raised and lowered with the $O(d,d)$ invariant metric $\eta_{mn}$. With this convention, the strong constraint reads $\eta^{mn} \partial_m \partial_n \diamond = 0$.

Now consider an extended tangent space with generalized vectors
\be
V^M  = (V^m , \sqrt{\alpha'}(t_m{}^n)_p{}^q \partial_q V^p)\, , \label{LorentzDiffOdd}
\ee
where the extended directions are not independent from the original ones and take values in the adjoint of $O(d,d)$. The $O(d,d)$ generators (which coincide with the $O(d,d)$ Killing metric)
\be
(t_p{}^q)_m{}^n = \frac 1 2 \left(\delta^n_p \delta^q_m - \eta^{qn} \eta_{pm}\right) = (t_m{}^n)_p{}^q\, ,
\ee
 can be used to define an invariant metric in the extended space
\be
\eta_{MP} = \left(\begin{matrix} \eta_{mp} & 0 \\ 0 & - 2 (t_m{}^n)_p{}^q\end{matrix}\right)\, .
\ee

The key observation is that the usual inner product and gauged generalized Lie derivative in the extended space
\bea
\langle\xi , V\rangle &=& \xi^M \eta_{MN} V^N\, , \nn\\
\left({\cal L}_\xi V\right)^M &=& \xi^P \partial_P V^M + (\eta^{MN} \eta_{PQ}\partial_N \xi^Q - \partial_P  \xi^M) V^P - f_{PQ}{}^M \xi^P V^Q\, , \label{ExtendedSpaceInnerProd}
\eea
where the gaugings correspond to the $O(d,d)$ gaugings in the extra directions, exactly reduce to the above equations (\ref{z1}) and (\ref{z2}) on the usual double tangent space
\bea
\langle\xi , V\rangle &=& \xi^m \eta_{mn} V^n - \alpha' \partial_m \xi^n \partial_n V^m\, ,\\
\left({\cal L}_\xi V\right)^m &=& \xi^p \partial_p V^m + (\eta^{mn} \eta_{pq}\partial_n \xi^q - \partial_p  \xi^m) V^p - \alpha' \eta^{mn}\partial_p V^q \partial_n \partial_q\xi^p\, ,
\eea
after implementing the strong constraint. In particular, when the constraint is solved in the frame in which everything depends only on the supergravity coordinates, one finds
\bea
\langle\xi , V\rangle &=& \xi^\mu V_\mu + \xi_\mu V^\mu - \alpha' \partial_\mu \xi^\rho \partial_\rho V^\mu\, ,\nn\\
\left({\cal L}_\xi V\right)^\mu &=& \xi^\rho \partial_\rho V^\mu - V^\rho \partial_\rho \xi^\mu\, , \label{gendiffHSZ}\\
\left({\cal L}_\xi V\right)_\mu &=& \xi^\rho \partial_\rho V_\mu + V_\rho \partial_\mu \xi^\rho - 2 V^\rho \partial_{[\rho} \xi_{\mu]} - \alpha' \partial_\rho V^\sigma \partial_\mu \partial_\sigma \xi^\rho\, .\nn
\eea

Then, we see that the $\alpha'$ corrections to the $O(d,d)$ inner product and generalized Lie derivative of \cite{DoubleAlpha} can be encoded in an extended space in which the inner product and generalized Lie derivative take the usual
expressions. This parallelism, while promising in order to reconcile both approaches, deserves a better understanding.

\section{Conclusions} \label{SECconclu}

In this paper we have extended the generalized flux formulation of DFT to include the
$O(\alpha ')$ corrections to the low energy effective heterotic string theory. This
includes the gauge and Lorentz Chern-Simons terms contained in the
field strength of the Kalb-Ramond field, as well as the
Yang-Mills and Riemann squared
 terms in the action. The gauge and Lorentz connections neatly fit together with the $d$-dimensional bein and two-form
field into an
enlarged generalized frame that transforms covariantly under a large duality group, part of which is gauged.
The Lorentz
connections included in the generalized frame can either be treated as dependent quantities
on the other fields or as independent connections which are finally related to the torsionful spin
 connection on-shell.
An important outcome
of this enhancement
is that the Riemann curvature tensor with torsion appears as one of
the components of the generalized fluxes. In this way,
being quadratic in fluxes,  the generalized action successfully
reproduces the
curvature squared term of the
heterotic effective theory. Hence, the construction allows to circumvent
 the issue raised in \cite{Riemann} about the absence
of a T-duality invariant four-derivative object, built from the
generalized metric, that reduces to the square of the Riemann tensor.

The gauging preserves a remnant $O(d,d)$ global symmetry that allowed us to compute the explicit  $\alpha '$ corrections to the Buscher rules. Indeed,
acting on the extended generalized metric
with factorized T-duality transformations,
we have found the first order $\alpha '$ corrections to the transformation rules of the massless bosonic heterotic fields.
These transformations serve as a
solution generating mechanism, as new solutions of the heterotic EOM
can be found by applying these rules to known solutions.

Several subsequent directions to extend these results suggest themselves.
One obvious course of future action is the construction of higher derivative terms. The ultimate goal is  to incorporate all order
$\alpha '$ corrections in a duality invariant formulation.
This is clearly a difficult problem and a more modest
target would be to understand these corrections order by order. Using duality
symmetries to determine higher derivative corrections
to supergravity has been a prolific area
of research in recent years (for example, see \cite{otros}
and references therein). It is possible
that higher order corrections require
 further enhancements of the duality group
and  additional extensions of the tangent space, in order to allow for
more degrees of freedom into a yet larger  generalized bein.

The supersymmetric extension is another direction of interest. Supersymmetric DFT was constructed in \cite{Siegel:1993th,Dan,SDFT} and more recently in gauged DFT in \cite{Berman:2013cli}. As explained in \cite{Bergshoeff:1989de}, the symmetry between the gauge and gravitational connections extends to the fermionic sector as well (more specifically the symmetry interchanges the gauginos with the curvature of the gravitinos), and this can be useful in the construction of the supersymmetric extension of our work.

It would also be interesting to explore $\alpha'$ corrections in the bosonic string and Type II superstring theories and see if they
 can be cast in a duality invariant form, similar to the one considered here.
One can already make contact with Type II theories by letting the heterotic gauge group be
embedded in the holonomy group. In this case, due to the symmetry between gauge and gravitational connections, the order $\alpha'$ terms in the action cancel each other,
in concordance with the fact that Type II theories only receive corrections of  order  $\alpha'^3$ and higher. The bosonic case will be discussed in a separate work \cite{bosonic}.

From a more phenomenological perspective,  compactifying this theory to lower dimensions
would allow to study  the quantum corrections to the low energy effective couplings and  scalar potential. Compactifications in manifolds with $SU(3)$ structure were performed in \cite{Grana:2014vva}, and it is also of interest to study supersymmetry preserving generalized Scherk-Schwarz compactifications along the lines of \cite{Aldazabal:2011nj} in this context.
The  deformations of the moduli space induced by $\alpha'$ corrections
may have important consequences in the search of  vacua
and the construction of sensible cosmological models in string theory. Moreover, it would allow to explore the relation between $\alpha'$ corrections and non-geometry, particularly the duality orbits of non-geometric fluxes discussed in \cite{Dibitetto:2012rk}, where the non-geometric effects are expected to be of order $\alpha'$.

~

\noindent{\bf Note 1.} At early stages of this work we
received a preliminary version of \cite{CGMW},
which contains some of the building blocks of our paper. This
includes the extended tangent space, inner product and generalized Lie
derivative. We would like to emphasize that a similar discussion on the relation
between this formalism and the one in  \cite{DoubleAlpha}  was first posed in \cite{CGMW}.

~

\noindent{\bf Note 2.} Soon after our work was posted, the papers \cite{GS} appeared, which
aim to describe bosonic and heterotic $\alpha'$ corrections following the approach in \cite{DoubleAlpha}.

~

\noindent{\bf Acknowledgments.} We are very  grateful to  M. Grana, the authors in \cite{CGMW},
A. Rosabal and  G. Thompson for valuable discussions. O. B. and
C. N.
 thank the Simons Foundation and the Abdus Salam I.C.T.P. for support and kind hospitality
during the final stages of this work.
Support by CONICET, UBA and ANPCyT  is also gratefully acknowledged.
\begin{appendix}

\section{Appendix}

\subsection{Conventions}

All through the paper we have used conventions that are useful to highlight the symmetry between the gauge and gravitational sectors. In this appendix we would like to discuss these conventions. Regarding the gauge sector, given the generators of the gauge group $t_\alpha$ we use the convention
\be
[t_\alpha,t_\beta] = -  f_{\alpha\beta}{}^\gamma t_\gamma \ , \ \ \ \ \ \kappa_{\alpha\beta} = tr (t_\alpha t_\beta)\, . \label{killinggauge}
 \ee
The gauge vectors are one forms in the adjoint of the gauge group $A_\mu = A_\mu{}^\alpha t_\alpha$ which is embedded in the fundamental of $G$.  We then have for example that
\bea
F_{\mu\nu} &=& 2 \partial_{[\mu} A_{\nu]} - 2 A_{[\mu} A_{\nu]}\, ,\\
F_{\mu\nu}{}^\alpha F_{\rho \sigma}{}^\beta \kappa_{\alpha\beta} &=& tr \left(F_{\mu\nu} F_{\rho\sigma}\right)\, ,\\
\partial_{[\mu }A_{\nu}{}^\alpha A_{\rho]}{}^\beta \kappa_{\alpha\beta} + \frac 1 3 f_{\alpha\beta\gamma} A_\mu{}^\alpha A_\nu{}^\beta A_\rho{}^\gamma
 &=& tr \left(A_{[\mu}\partial_{\nu} A_{\rho]}  - \frac 2 3 A_{[\mu} A_\nu A_{\rho]}\right)\, .\eea
Similarly, given the generators of the Lorentz group $t_\Lambda$ we use the convention
\be
[t_\Lambda, t_\Gamma] = -  f_{\Lambda\Gamma}{}^\Sigma t_\Sigma \ , \ \ \ \ \ \kappa_{\Lambda\Gamma} = - tr(t_\Lambda t_{\Gamma})\, . \label{killinglorentz}
\ee
The spin connection (with torsion) is a one-form in the adjoint of the Lorentz group $\tilde \omega_\mu = \tilde\omega_\mu{}^\Lambda t_\Lambda$ which is embedded in the fundamental of $G$ as well. For this sector we then find
\bea
\tilde R_{\mu\nu} &=& 2 \partial_{[\mu} \tilde \omega_{\nu]} - 2 \tilde \omega_{[\mu} \tilde\omega_{\nu]}\, ,\\
\tilde R_{\mu\nu}{}^\Lambda \tilde R_{\rho \sigma}{}^\Gamma \kappa_{\Lambda\Gamma} &=& - tr \left(\tilde R_{\mu\nu} \tilde R_{\rho\sigma}\right)\, ,\\
\partial_{[\mu }\tilde \omega_{\nu}{}^\Lambda \tilde \omega_{\rho]}{}^\Gamma \kappa_{\Lambda\Gamma} + \frac 1 3 f_{\Lambda\Gamma\Sigma} \tilde \omega_\mu{}^\Lambda \tilde \omega_\nu{}^\Gamma \tilde\omega_\rho{}^\Sigma
 &=& - tr \left(\tilde\omega_{[\mu}\partial_{\nu} \tilde\omega_{\rho]}  - \frac 2 3 \tilde\omega_{[\mu} \tilde\omega_\nu \tilde\omega_{\rho]}\right)\, .
\eea
Note the different conventions used for the Killing metrics in the gauge (\ref{killinggauge}) and Lorentz (\ref{killinglorentz}) sectors.

\subsection{Comments on the equations of motion}

In this Appendix we outline the procedure to obtain the equations of motion to $O(\alpha')$, $i.e.$ (\ref{DeltaPhi})-(\ref{eqA}). For further details we refer to \cite{Bergshoeff:1989de}-\cite{Becker:2009zx} and references therein. Specifically we only focus on the EOM for the two-form and the bein, since their derivation is subtle due the fact that they are both implicitly contained in the torsionful spin connection.

We consider first the equation of motion for the two-form. Explicitly, the Lagrangian does not depend on $B_{\mu\nu}$, but on its derivatives. Implicitly, it depends on first and second derivatives of $B_{\mu\nu}$ through the torsionful spin connection and its derivatives.  Therefore, the full variation of the action is
\bea
\delta_B \nonumber  S &=& \int dx \left[ \frac{\delta {\cal L}}{\delta \partial_\lambda B_{\rho\nu}}\ \delta \partial_\lambda B_{\rho\nu} + \frac {\delta {\cal L}}{\delta \omega^{(-)}_{\mu\bar a}{}^{\bar b}}\ \frac{\delta \omega^{(-)}_{\mu\bar a}{}^{\bar b}}{\delta \partial_\lambda B_{\rho\nu}}\ \delta \partial_\lambda B_{\rho\nu} \right. \\ \nonumber
&&~~~~~~~~~\left. +\ \frac{\delta {\cal L}}{\delta \partial_\eta \omega^{(-)}_{\mu\bar a}{}^{\bar b}}\left(\frac{\delta \partial_\eta \omega^{(-)}_{\mu\bar a}{}^{\bar b}}{\delta \partial_\lambda B_{\rho\nu}}\ \delta \partial_\lambda B_{\rho\nu}+
\frac{\delta \partial_\eta \omega^{(-)}_{\mu\bar a}{}^{\bar b}}{\delta \partial_\xi\partial_\lambda B_{\rho\nu}}\ \delta \partial_\xi\partial_\lambda B_{\rho\nu}\right) \right]\, ,
\eea
which can be rewritten, after integrating by parts, as
\be \label{EOMB}\delta_B S=  \int dx \left[-\partial_\lambda \frac{\delta {\cal L}}{\delta \partial_\lambda B_{\rho\nu}} \delta  B_{\rho\nu} + \left(\frac {\delta {\cal L}}{\delta \omega^{(-)}_{\mu\bar a}{}^{\bar b}} - \partial_\eta \frac{\delta {\cal L}}{\delta \partial_\eta \omega^{(-)}_{\mu\bar a}{}^{\bar b}}\right) \delta_B \omega^{(-)}_{\mu\bar a}{}^{\bar b} \right].
\ee
It is straightforward to compute the expression multiplying
$\delta_B \omega^{(-)}$ in (\ref{EOMB}), that we
 denote $\delta_{\omega^{(-)}}{\cal L}$,
and the result is
\be \label{delomL}
\delta_{\omega^{(-)}} {\cal L} = \frac {\alpha'} 2 \sqrt{-g}\nabla_\rho \left( e^{-2\phi} H^{\rho \mu\nu}\right)\omega^{(-)}_{\nu\bar b}{}^{\bar a} + \alpha' \sqrt{-g} \nabla_\nu^{(+,-)}\left(e^{-2\phi} R^{(-) \mu\nu}{}_{\bar b}{}^{\bar a}\right)\, .
\ee
Using
\be \label{delom}\delta_B \omega^{(-)}_{\mu \bar a}{}^{\bar b}  = \frac{\delta \omega^{(-)}_{\mu \bar a}{}^{\bar b}}{\delta \partial_\lambda B_{\rho\nu}}\delta \partial_\lambda B_{\rho\nu} = -3 \delta_\mu ^{[\lambda}e_{\bar a}{}^\rho e_{\bar c}{}^{\nu ]}s^{\bar c\bar b} \delta \partial_\lambda B_{\rho\nu}  + O(\alpha')\, , \ee
replacing in (\ref{EOMB}) and integrating by parts, we get
\bea \delta_B {\cal L} &=& \sqrt{-g}\ \nabla_\lambda \left(e^{-2\Phi} H^{\lambda\rho\nu}\right) \delta B_{\rho\nu} -3
\alpha' \partial_\lambda \left(\frac{\sqrt{-g}}{2} \nabla_\eta (e^{-2\Phi} H^{\eta\mu[\lambda})\omega_{\mu}{}^{\nu\rho]} \right. \\ \nonumber
&&\left. +\ \sqrt{-g}\  \nabla_\mu ^{(+,-)} \left(e^{-\Phi} {\cal R}^{\mu[\lambda\nu\rho]}\right)\vphantom{\frac{\sqrt{-g}}{2}}\right) \delta B_{\rho\nu} + O(\alpha'^2)\, .\eea
Here, the first term is what we denoted $\Delta B$  in (\ref{DeltaB}) and
the second one is proportional to both $\Delta B$ and $\Delta g$ defined in
(\ref{Deltag})  (see for example  (\ref{SolExtraEOM})). This is a very useful and well known result obtained in \cite{Bergshoeff:1989de}. However, because of the complete antisymmetry in the indices $\lambda$, $\rho$, $\nu$, the dependence on $\Delta g$ (which is symmetric) vanishes. One ends with
\be \label{eomb}\delta_B {\cal L} =
\sqrt{-g}\left(\Delta B^{\rho\nu}  +
\alpha' \hat{\cal O}^{\rho\nu}[\Delta B] + O(\alpha'^2)
\right)\delta B_{\rho\nu}\, ,\ee
where
the linear differential operator is defined by
\be
\hat{\cal O}^{\rho\nu}[\Delta B]
= \frac{3}{2\sqrt{-g}} \partial_\lambda \Big(\sqrt{-g}\Delta B^{\mu[\lambda} \omega^{(-)}_{\mu}{}^{\rho\nu]} + \sqrt{-g} g^{\mu [\lambda} \nabla_\mu
^{(+)}\Delta B^{\rho\nu ]} \Big) \ .
\ee
We then see from (\ref{eomb}) that the
 EOM for the $B$-field takes the following schematic form
\be \label{eombfinal}
\left(1 + \alpha' \hat {\cal O}\right) \Delta B = O(\alpha'^2)\, .
\ee
Remarkably,
since the operator $1 + \alpha' \hat {\cal O}$ is invertible
to $O(\alpha')$, one ends with
\be
\Delta B_{\mu\nu} = O(\alpha'^2)\ .
\ee

Using this equation and implementing the same strategy for the EOM of the bein,
one finds an identical equation to (\ref{eombfinal}) with a different operator
\be
\left(1 + \alpha' \tilde {\cal O}\right) \Delta g = O(\alpha'^2)\ .
\ee
Again, the details of this operator are not important in this calculation, and the only thing that matters is that to $O(\alpha')$ the operator $1 + \alpha' \tilde {\cal O}$ is invertible, so that one finally obtains
\be
\Delta g_{\mu\nu} = O(\alpha'^2)\ .
\ee

We then see that the EOM for the bein and two-form, properly computed from the full variation of the action, are equivalent (to $O(\alpha')$) to the EOM that one would obtain by only varying the action with respect to the explicit dependence of the fields, and treating the torsionful spin connection as an independent field.
\end{appendix}

\label{sec:References}


\begin{thebibliography}{98}

\bibitem{Siegel:1993th}
  W.~Siegel,
  ``Superspace duality in low-energy superstrings,''
  Phys.\ Rev.\ D {\bf 48} (1993) 2826
  [hep-th/9305073].

  W.~Siegel,
  ``Two vierbein formalism for string inspired axionic gravity,''
  Phys.\ Rev.\ D {\bf 47} (1993) 5453
  [hep-th/9302036].

 W.~Siegel,
  ``Manifest duality in low-energy superstrings,''
  In *Berkeley 1993, Proceedings, Strings '93* 353-363, and State U. New York Stony Brook - ITP-SB-93-050 (93,rec.Sep.) 11 p. (315661)
  [hep-th/9308133].

\bibitem{Hull:2009mi}
  C.~Hull and B.~Zwiebach,
  ``Double Field Theory,''
  JHEP {\bf 0909} (2009) 099
  [arXiv:0904.4664 [hep-th]].

    O.~Hohm, C.~Hull and B.~Zwiebach,
  ``Background independent action for double field theory,''
  JHEP {\bf 1007} (2010) 016
  [arXiv:1003.5027 [hep-th]].




\bibitem{heteroticHohm}

 O.~Hohm and S.~K.~Kwak,
  ``Double Field Theory Formulation of Heterotic Strings,''
  JHEP {\bf 1106} (2011) 096
  [arXiv:1103.2136 [hep-th]].


  \bibitem{TypeII}

   O.~Hohm, S.~K.~Kwak and B.~Zwiebach,
  ``Double Field Theory of Type II Strings,''
  JHEP {\bf 1109} (2011) 013
  [arXiv:1107.0008 [hep-th]].

  O.~Hohm and S.~K.~Kwak,
  ``Massive Type II in Double Field Theory,''
  JHEP {\bf 1111} (2011) 086
  [arXiv:1108.4937 [hep-th]].


  I.~Jeon, K.~Lee, J.~-H.~Park and Y.~Suh,
  ``Stringy Unification of Type IIA and IIB Supergravities under N=2 D=10
Supersymmetric Double Field Theory,''
  arXiv:1210.5078 [hep-th].

\bibitem{Dan}


   A.~Coimbra, C.~Strickland-Constable and D.~Waldram,
  ``Supergravity as Generalised Geometry I: Type II Theories,''
  JHEP {\bf 1111} (2011) 091
  [arXiv:1107.1733 [hep-th]].

 \bibitem{SDFT}
   O.~Hohm and S.~K.~Kwak,
  ``N=1 Supersymmetric Double Field Theory,''
  JHEP {\bf 1203} (2012) 080
  [arXiv:1111.7293 [hep-th]].

  I.~Jeon, K.~Lee and J.~-H.~Park,
  ``Supersymmetric Double Field Theory: Stringy Reformulation of Supergravity,''
  Phys.\ Rev.\ D {\bf 85} (2012) 081501
   [Erratum-ibid.\ D {\bf 86} (2012) 089903]
  [arXiv:1112.0069 [hep-th]].




  \bibitem{Generalizedmetric}

  O.~Hohm, C.~Hull and B.~Zwiebach,
  ``Generalized metric formulation of double field theory,''
  JHEP {\bf 1008}, 008 (2010)
  [arXiv:1006.4823 [hep-th]].

\bibitem{framelikegeom}

O.~Hohm and S.~K.~Kwak,
  ``Frame-like Geometry of Double Field Theory,''
  J.\ Phys.\ A {\bf 44} (2011) 085404
  [arXiv:1011.4101 [hep-th]].

  \bibitem{genLie}
  M.~Grana, R.~Minasian, M.~Petrini and D.~Waldram,
  ``T-duality, Generalized Geometry and Non-Geometric Backgrounds,''
  JHEP {\bf 0904} (2009) 075
  [arXiv:0807.4527 [hep-th]].

  \bibitem{Cbracket}
  C.~Hull and B.~Zwiebach,
  ``The Gauge algebra of double field theory and Courant brackets,''
  JHEP {\bf 0909} (2009) 090
  [arXiv:0908.1792 [hep-th]].




\bibitem{Reviews}
G.~Aldazabal, D.~Marques and C.~Nunez,
  ``Double Field Theory: A Pedagogical Review,''
  Class.\ Quant.\ Grav.\  {\bf 30} (2013) 163001
  [arXiv:1305.1907 [hep-th]].

  D.~S.~Berman and D.~C.~Thompson,
  ``Duality Symmetric String and M-Theory,''
  arXiv:1306.2643 [hep-th].

  \bibitem{DoubleAlpha}
  O.~Hohm, W.~Siegel and B.~Zwiebach,
  ``Doubled $\alpha'$-geometry,''
  JHEP {\bf 1402} (2014) 065
  [arXiv:1306.2970 [hep-th]].




\bibitem{cai}
  Y.~Cai and C.~A.~Nunez,
  ``Heterotic String Covariant Amplitudes and Low-energy Effective Action,''
  Nucl.\ Phys.\ B {\bf 287}, 279 (1987).

 D.~J.~Gross and J.~H.~Sloan,
  ``The Quartic Effective Action for the Heterotic String,''
  Nucl.\ Phys.\ B {\bf 291}, 41 (1987).

\bibitem{Bergshoeff:1988nn}
  E.~Bergshoeff and M.~de Roo,
  ``Supersymmetric Chern-Simons Terms in Ten-dimensions,''
  Phys.\ Lett.\ B {\bf 218}, 210 (1989).

 \bibitem{Bergshoeff:1989de}
  E.~A.~Bergshoeff and M.~de Roo,
  ``The Quartic Effective Action of the Heterotic String and Supersymmetry,''
  Nucl.\ Phys.\ B {\bf 328} (1989) 439.

\bibitem{Callan:1985ia}
  C.~G.~Callan, Jr., D.~Friedan, E.~J.~Martinec and M.~J.~Perry,
  ``Strings in Background Fields,''
  Nucl.\ Phys.\ B {\bf 262} (1985) 593.

  \bibitem{Metsaev:1987zx}
  R.~R.~Metsaev and A.~A.~Tseytlin,
  ``Order alpha-prime (Two Loop) Equivalence of the String Equations of Motion and the Sigma Model Weyl Invariance
Conditions: Dependence on the Dilaton and the Antisymmetric Tensor,''
  Nucl.\ Phys.\ B {\bf 293} (1987) 385.

  \bibitem{Chemissany:2007he}
  W.~A.~Chemissany, M.~de Roo and S.~Panda,
  ``alpha'-Corrections to Heterotic Superstring Effective Action Revisited,''
  JHEP {\bf 0708} (2007) 037
  [arXiv:0706.3636 [hep-th]].

\bibitem{Becker:2009df}
  K.~Becker and S.~Sethi,
  ``Torsional Heterotic Geometries,''
  Nucl.\ Phys.\ B {\bf 820}, 1 (2009)
  [arXiv:0903.3769 [hep-th]].

\bibitem{Becker:2009zx}
  K.~Becker, C.~Bertinato, Y.~C.~Chung and G.~Guo,
  ``Supersymmetry breaking, heterotic strings and fluxes,''
  Nucl.\ Phys.\ B {\bf 823} (2009) 428
  [arXiv:0904.2932 [hep-th]].

\bibitem{dualt1}
  M.~J.~Duff, B.~E.~W.~Nilsson, N.~P.~Warner and C.~N.~Pope,
  ``{Kaluza-Klein} Approach to the Heterotic String. 2.,''
  Phys.\ Lett.\ B {\bf 171} (1986) 170.

   \bibitem{dualt2}
   D.~Andriot,
  ``Heterotic string from a higher dimensional perspective,''
  Nucl.\ Phys.\ B {\bf 855} (2012) 222
  [arXiv:1102.1434 [hep-th]].

  \bibitem{math1}

  M.~Garcia-Fernandez,
  ``Torsion-free generalized connections and Heterotic Supergravity,''
  arXiv:1304.4294 [math.DG].

  \bibitem{math2}

  D.~Baraglia and P.~Hekmati,
  ``Transitive Courant Algebroids, String Structures and T-duality,''
  arXiv:1308.5159 [math.DG].
\bibitem{CGMW}
 A.~Coimbra, R.~Minasian, H.~Triendl and D.~Waldram,
  ``Generalised geometry for string corrections,''
  arXiv:1407.7542 [hep-th].

\bibitem{Exploring}
D.~Geissbuhler, D.~Marques, C.~Nunez and V.~Penas,
  ``Exploring Double Field Theory,''
  JHEP {\bf 1306} (2013) 101
  [arXiv:1304.1472 [hep-th]].



\bibitem{NaturalCurvature}
  M.~Polacek and W.~Siegel,
  ``Natural curvature for manifest T-duality,''
  JHEP {\bf 1401} (2014) 026
  [arXiv:1308.6350 [hep-th]].

M.~Hatsuda, K.~Kamimura and W.~Siegel,
  ``Superspace with manifest T-duality from type II superstring,''
  arXiv:1403.3887 [hep-th].

    M.~Polacek and W.~Siegel,
  ``T-duality off shell in 3D Type II superspace,''
  JHEP {\bf 1406} (2014) 107
  [arXiv:1403.6904 [hep-th]].

\bibitem{Riemann}
  O.~Hohm and B.~Zwiebach,
  ``On the Riemann Tensor in Double Field Theory,''
  JHEP {\bf 1205} (2012) 126
  [arXiv:1112.5296 [hep-th]].

  \bibitem{buscher1}
T. Buscher, ``A Symmetry of the String Background Field Equations,'' Phys. Lett.
{\bf 194B} (1987) 59.

T. Buscher, ``Path Integral Derivation of Quantum Duality in
Nonlinear Sigma Models,''
Phys. Lett. {\bf 201B} (1988) 466.

\bibitem{Rocek:1991ps}
  M.~Rocek and E.~P.~Verlinde,
  ``Duality, quotients, and currents,''
  Nucl.\ Phys.\ B {\bf 373}, 630 (1992)
  [hep-th/9110053].

\bibitem{Alvarez:1994wj}
  E.~Alvarez, L.~Alvarez-Gaume and Y.~Lozano,
  ``A Canonical approach to duality transformations,''
  Phys.\ Lett.\ B {\bf 336}, 183 (1994)
  [hep-th/9406206].


  E.~Alvarez, L.~Alvarez-Gaume and Y.~Lozano,
  ``An Introduction to T duality in string theory,''
  Nucl.\ Phys.\ Proc.\ Suppl.\  {\bf 41}, 1 (1995)
  [hep-th/9410237].

\bibitem{qbusch}

E.~Bergshoeff, I.~Entrop and R.~Kallosh,
  ``Exact duality in string effective action,''
  Phys.\ Rev.\ D {\bf 49}, 6663 (1994)
  [hep-th/9401025].

  K.~A.~Meissner,
  ``Symmetries of higher order string gravity actions,''
  Phys.\ Lett.\ B {\bf 392} (1997) 298
  [hep-th/9610131].

 N.~Kaloper and K.~A.~Meissner,
  ``Duality beyond the first loop,''
  Phys.\ Rev.\ D {\bf 56}, 7940 (1997)
  [hep-th/9705193].

J.~Balog, P.~Forgacs, N.~Mohammedi, L.~Palla and J.~Schnittger,
  ``On quantum T duality in sigma models,''
  Nucl.\ Phys.\ B {\bf 535}, 461 (1998)
  [hep-th/9806068].

M.~Serone and M.~Trapletti,
  ``A Note on T-duality in heterotic string theory,''
  Phys.\ Lett.\ B {\bf 637}, 331 (2006)
  [hep-th/0512272].

\bibitem{bergortin}
E.~Bergshoeff, B.~Janssen and T.~Ortin,
  ``Solution generating transformations and the string effective action,''
  Class.\ Quant.\ Grav.\  {\bf 13}, 321 (1996)
  [hep-th/9506156].



  \bibitem{GDFT}
  M.~Grana and D.~Marques,
  ``Gauged Double Field Theory,''
  JHEP {\bf 1204} (2012) 020
  [arXiv:1201.2924 [hep-th]].

  \bibitem{Aldazabal:2011nj}
  G.~Aldazabal, W.~Baron, D.~Marques and C.~Nunez,
  ``The effective action of Double Field Theory,''
  JHEP {\bf 1111} (2011) 052
   [Erratum-ibid.\  {\bf 1111} (2011) 109]
  [arXiv:1109.0290 [hep-th]].

  D.~Geissbuhler,
  ``Double Field Theory and N=4 Gauged Supergravity,''
  JHEP {\bf 1111} (2011) 116
  [arXiv:1109.4280 [hep-th]].

\bibitem{Lee:2014mla}
  K.~Lee, C.~Strickland-Constable and D.~Waldram,
  ``Spheres, generalised parallelisability and consistent truncations,''
  arXiv:1401.3360 [hep-th].


\bibitem{Green:1984sg}
  M.~B.~Green and J.~H.~Schwarz,
  ``Anomaly Cancellation in Supersymmetric D=10 Gauge Theory and Superstring Theory,''
  Phys.\ Lett.\ B {\bf 149} (1984) 117.

  \bibitem{Stringydiffgeom}
  I.~Jeon, K.~Lee and J.~-H.~Park,
  ``Differential geometry with a projection: Application to double
field theory,''
  JHEP {\bf 1104} (2011) 014
  [arXiv:1011.1324 [hep-th]].

    I.~Jeon, K.~Lee and J.~-H.~Park,
  ``Stringy differential geometry, beyond Riemann,''
  Phys.\ Rev.\ D {\bf 84} (2011) 044022
  [arXiv:1105.6294 [hep-th]].



\bibitem{Giveon:1994fu}
  A.~Giveon, M.~Porrati and E.~Rabinovici,
  ``Target space duality in string theory,''
  Phys.\ Rept.\  {\bf 244} (1994) 77
  [hep-th/9401139].

\bibitem{otros}
 M.~B.~Green, S.~D.~Miller, J.~G.~Russo and P.~Vanhove,
  ``Eisenstein series for higher-rank groups and string theory amplitudes,''
  Commun.\ Num.\ Theor.\ Phys.\  {\bf 4}, 551 (2010)
  [arXiv:1004.0163 [hep-th]].

M.~B.~Green, J.~G.~Russo and P.~Vanhove,
  ``String theory dualities and supergravity divergences,''
  JHEP {\bf 1006}, 075 (2010)
  [arXiv:1002.3805 [hep-th]].

M. R. Garousi, ``T-duality of the Riemann curvature corr
ections to supergravity,''
Phys.Lett.
B718
(2013) 1481,
arXiv:1208.4459 [hep-th]

J. T. Liu and R. Minasian, ``Higher-derivative coupling
s in string theory: dualities and the
B-field,''
arXiv:1304.3137 [hep-th]

  H.~Godazgar and M.~Godazgar,
  ``Duality completion of higher derivative corrections,''
  JHEP {\bf 1309} (2013) 140
  [arXiv:1306.4918 [hep-th]].


 \bibitem{Berman:2013cli}
  D.~S.~Berman and K.~Lee,
  ``Supersymmetry for Gauged Double Field Theory and Generalised Scherk-Schwarz Reductions,''
  Nucl.\ Phys.\ B {\bf 881} (2014) 369
  [arXiv:1305.2747 [hep-th]].



\bibitem{bosonic}
O.~Bedoya, D.~Marques and C.~Nu\~nez, in preparation.

\bibitem{Grana:2014vva}
  M.~Grana, J.~Louis, U.~Theis and D.~Waldram,
  ``Quantum Corrections in String Compactifications on SU(3) Structure Geometries,''
  arXiv:1406.0958 [hep-th].

  \bibitem{Dibitetto:2012rk}
  G.~Dibitetto, J.~J.~Fernandez-Melgarejo, D.~Marques and D.~Roest,
  ``Duality orbits of non-geometric fluxes,''
  Fortsch.\ Phys.\  {\bf 60} (2012) 1123
  [arXiv:1203.6562 [hep-th]].




\bibitem{GS}
 O.~Hohm and B.~Zwiebach,
  ``Green-Schwarz mechanism and $\alpha'$-deformed Courant brackets,''
  arXiv:1407.0708 [hep-th].

   O.~Hohm and B.~Zwiebach,
  ``Double Field Theory at Order $\alpha'$,''
  arXiv:1407.3803 [hep-th].

\end{thebibliography}
\end{document}